\documentclass[sigconf]{acmart}

\usepackage{graphicx}
\usepackage{multirow}
\usepackage{caption}
\usepackage{subcaption}
\usepackage{graphicx}
\usepackage{booktabs}
\usepackage{caption}
\usepackage{siunitx}
\usepackage{placeins}
\usepackage{float}
\newcommand{\new}[1]{{\textcolor{black}{#1}}}

\AtBeginDocument{%
  }

\setcopyright{acmlicensed}

\copyrightyear{2025} \acmYear{2025} \setcopyright{cc} \setcctype{by} \acmConference[FAccT '25]{The 2025 ACM Conference on Fairness, Accountability, and Transparency}{June 23--26, 2025}{Athens, Greece} \acmBooktitle{The 2025 ACM Conference on Fairness, Accountability, andTransparency (FAccT '25), June 23--26, 2025, Athens, Greece}

\acmDOI{10.1145/3715275.3732123} 
\acmISBN{979-8-4007-1482-5/2025/06}

\begin{document}

\title{Bias Delayed is Bias Denied? Assessing the Effect of Reporting Delays on Disparity Assessments}

\author{Jennah Gosciak}
\authornote{Both authors contributed equally to this research.}
\email{jrg377@cornell.edu}
\affiliation{\institution{Cornell University}
\city{Ithaca}
\state{NY}
\country{USA}}

\author{Aparna Balagopalan}
\authornotemark[1]
\email{aparnab@mit.edu}
\affiliation{\institution{Massachusetts Institute of Technology}
\city{Cambridge}
\state{MA}
\country{USA}}

\author{Derek Ouyang}
\affiliation{\institution{Stanford University}
\city{Stanford}
\state{CA}
\country{USA}}

\author{Allison Koenecke}
\affiliation{\institution{Cornell University}
\city{Ithaca}
\state{NY}
\country{USA}}

\author{Marzyeh Ghassemi}
\affiliation{\institution{Massachusetts Institute of Technology}
\city{Cambridge}
\state{MA}
\country{USA}}

\author{Daniel E. Ho}
\affiliation{\institution{Stanford University}
\city{Stanford}
\state{CA}
\country{USA}}

\renewcommand{\shortauthors}{Gosciak et al.}

\begin{abstract}
  Prior work has documented widespread racial and ethnic inequities across sectors, such as healthcare, finance, and technology. Across all of these domains, conducting disparity assessments at regular time intervals is critical for surfacing potential biases in decision-making and improving outcomes across demographic groups. Because disparity assessments fundamentally depend on the availability of demographic information, their efficacy is limited by the availability and consistency of available demographic identifiers. While prior work has considered the impact of \emph{missing} data on fairness, little attention has been paid to the role of \emph{delayed} demographic data. Delayed data, while eventually observed, might be missing at the critical point of monitoring and action -- and delays may be unequally distributed across groups in ways that distort disparity assessments. We characterize such impacts in healthcare, using electronic health records of over 5M patients across primary care practices in all 50 states. Our contributions are threefold.  First, we document the high rate of race and ethnicity reporting delays in a healthcare setting and demonstrate widespread variation in rates at which demographics are reported across different groups. Second, through a set of retrospective analyses using real data, we find that such delays impact disparity assessments and hence conclusions made across a range of consequential healthcare outcomes, particularly at more granular levels of state-level and practice-level assessments. Third, we find limited ability of conventional methods that impute missing race in mitigating the effects of reporting delays on the accuracy of timely disparity assessments. Our insights and methods generalize to many domains of algorithmic fairness where delays in the availability of sensitive information may confound audits, thus deserving closer attention within a pipeline-aware machine learning framework.
\end{abstract}




\maketitle
\section{Introduction}

Racial inequity in the United States (U.S.) remains a significant issue in sectors such as healthcare, employment, finance, and education.\footnote{For brevity, we use ``race'' to refer to both race and ethnicity throughout the remainder of this paper.} %
In healthcare, where such disparities can be stark \citep{howell2018reducing,haw2021diabetes,allen2010racial}, %
researchers, policymakers, and healthcare institutions have increasingly turned toward assessments to measure, and potentially mitigate, such disparities~\cite{hemrp_ca}. Such assessments are also crucial tools for auditing the fairness of machine learning (ML)-based diagnostic tools --- an area of growing concern as ML and data-driven decision-making become more prominent in healthcare~\cite{beaulieu2019trends}. %

Less recognized is a core impediment for disparity assessments: the \textit{timely reporting of demographic information} (\textit{e.g.}, race, gender) by patients and providers. In this work, we show that failing to account for reporting delays, as distinct from missing data, can obfuscate health disparities. Leveraging access to a large, longitudinal dataset of over 5M patients, sourced from primary care practices throughout the U.S., we both document the extent of race reporting delays and examine the effect on disparity assessments, which we expect to be increasingly common. %

\begin{figure*}
\centering
\includegraphics[width=\linewidth,trim={3cm 5.5cm 3cm 4.5cm},clip]{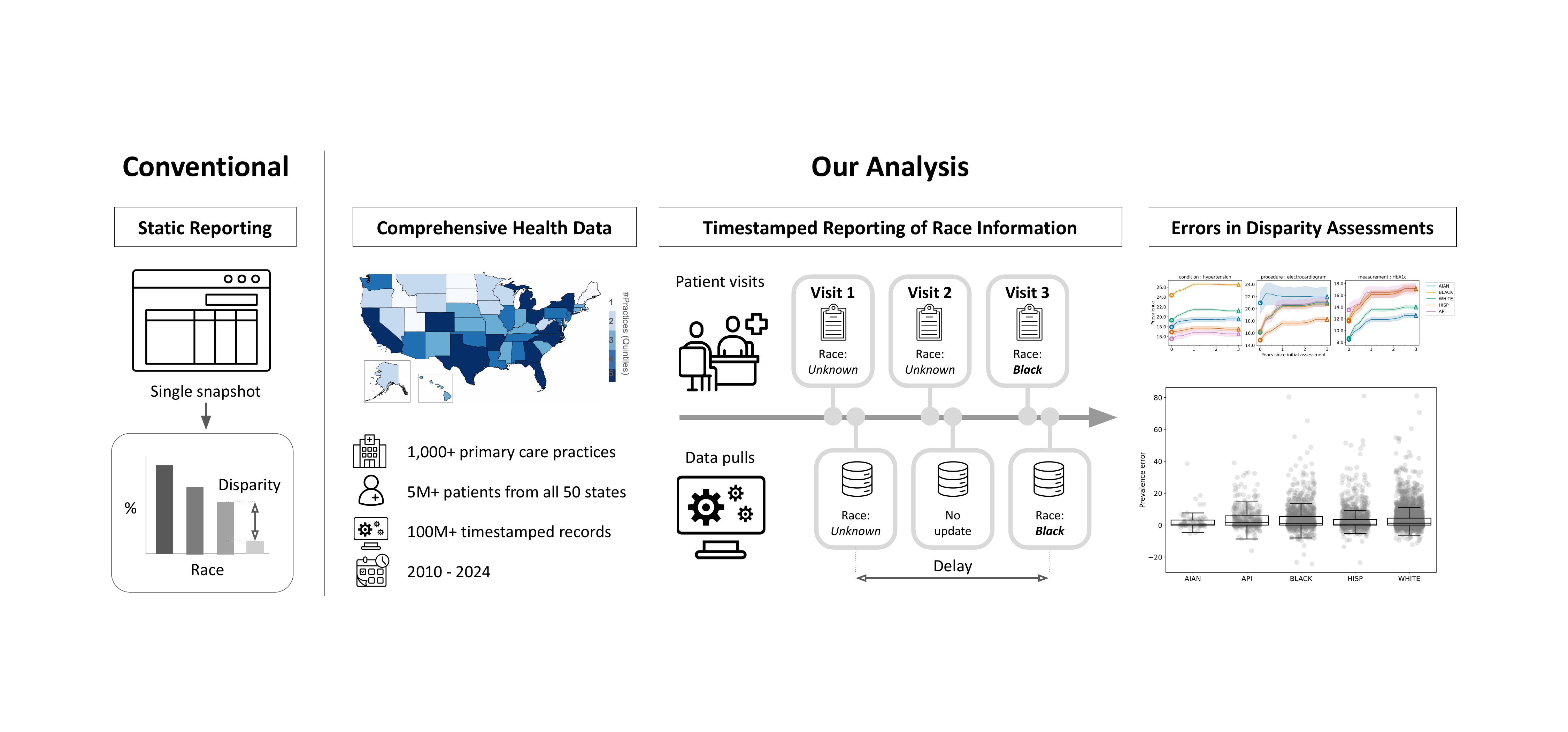}
\caption{On the left, we show the \emph{conventional} approach to disparity assessments, which results in a \emph{static} measure of disparity. On the right, we present the three core components of our analysis. First, we leverage access to a \emph{comprehensive} dataset of over 1,000 primary care practices, 5M patients from all 50 states, and 100M patient interactions from 2010 to 2024. Second, we use \emph{timestamped} records to identify and measure delays in reporting of race information. Third, we demonstrate how reporting delays drive \emph{errors} in disparity assessments across a variety of consequential health outcomes, from the national (Figure~\ref{fig:simulation}) to the practice level (Figure~\ref{fig:practice-dist-differences}).}
\Description{Diagram that contrasts a dynamic approach to assessing disparities considering delays over multiple time points with a conventional, static approach. On the left, the conventional approach is shown with two illustrations --- a snapshot of a tabular dataset and a bar graph representing a disparity assessment conducted at one point in time. On the right, three different characteristics of our analysis are shown: (1) a map of the full AFC dataset to depict its comprehensive coverage, a schematic representing patient visits and data pulls over time, and (3) figures from the results of this analysis.}
\label{fig:cohort-schematic}
\end{figure*}

We make several contributions in this work. First, we provide researchers and practitioners with a concrete definition of \emph{delay}, which occurs when information is initially unreported for an individual, but eventually becomes available after repeated interactions with data collection systems (\textit{e.g.}, repeated patient visits to a primary care provider). Importantly, such delays may affect variables that are considered ``static''~\cite{caballero2015dynamically} (\textit{e.g.}, data usually collected at the time of hospital admission such as race and pre-existing diagnoses~\cite{li2021integrating}).
No prior works in healthcare or fair machine learning, to the best of our knowledge, have rigorously analyzed the impact of this type of temporal missingness of demographic attributes in administrative data, which we are able to observe through a richly timestamped healthcare dataset. We show that, in fact, delays are widespread. Second, we examine heterogeneity in reporting delays and find that rates of delayed reporting vary by race (and other healthcare attributes), directly implicating bias concerns. Third, we design and carry out a series of retrospective analyses on this data to understand how delayed race reporting impacts disparity assessments in a real-world, high-impact setting. We find consequential distortions, with prevalence errors of 10 percentage points or more not uncommon at the practice level. Lastly, we demonstrate that widely used imputation methods like Bayesian Improved First Name Surname Geocoding (BIFSG)~\cite{voicu2018using}, while relatively accurate at individual prediction of race, do not significantly reduce errors in disparity assessment across all outcomes of interest.

Our work highlights the importance of pipeline-aware, context-specific approaches to data-driven decision making~\citep{black2023toward, suresh2021framework, akpinar2022sandbox}.  Pipeline-aware fairness involves considering all of the different design decisions in the full ML pipeline and their effect on fairness outcomes. As \citet{black2023toward} demonstrate, far more effort  has been spent studying bias in statistical models. Much less attention has been paid to other aspects of the ML pipeline such as data collection -- the focus of our paper.
In settings involving time-sensitive, routine disparity assessments (\textit{e.g.}, dashboards measuring health outcomes for different racial groups), delayed race data may hinder \textit{responsive} and \textit{actionable} feedback. Our results suggest researchers and practitioners should expend greater efforts to identify sources of delay that might exist within real-world data collection pipelines, consider their downstream impacts, and test policy and/or programmatic interventions to reduce delays. As we show, delayed reporting may lead to inaccurate and misleading estimates of disparities, with direct fairness implications; these findings may similarly affect other high-stakes applications and geographic domains. Figure~\ref{fig:cohort-schematic} summarizes the value of our analytic approach, which surfaces errors in disparity assessments across time and geography by leveraging timestamped race reporting information from a unique health dataset, all of which would not be possible via a conventional approach to disparity assessments using more static data.

The rest of the paper is structured as follows. In Section~\ref{sec:background}, we discuss the changing policy landscape related to monitoring and addressing health disparities in the U.S. We also connect our work to practical challenges in algorithmic fairness in the wild: \textit{(1)} the frequency of missing demographic data in many real-world contexts, and \textit{(2)} the importance of studying fairness \emph{dynamically}. Section~\ref{sec:data} provides an overview of our dataset, which uniquely affords us access to information from over 1,000 practices across all 50 states and over 5M patients in the U.S. Most notably, the data contains fine-grained longitudinal information across over 100M patient interactions, including timestamped reporting of race information, which enables us to design realistic assessments of the magnitude, correlates, and impact of reporting delays. Then, in Section~\ref{sec:methods} we detail our methods, including details on data processing, key definitions, and summary metrics. Section~\ref{sec:delayed-reporting-desc} describes the population of patients with and without delays and presents the results from our retrospective analyses conducted on real patient data. 
Lastly, in Section~\ref{sec:discussion}, we discuss the implications of our findings for researchers and practitioners in both healthcare and algorithmic fairness. In particular, we call attention to the importance of considering fairness in real-world deployment settings where the reporting mechanism for demographic attributes may lead to delays over time.\looseness=-1

\section{Background and Related Work}
\label{sec:background}
Prior research has extensively documented racial health disparities in the U.S., from pain management to life expectancy \citep{green2003unequal, anderson2009racial, baciu2017communities, hauck2011racial, macdorman2021racial, dwyer2022life}. However, there are several impediments to \emph{accurate and timely assessments} of disparities. In our work, we focus on one challenge that has been neglected in prior work, but is of immense practical consequence: reporting delays in demographic information. While several prior works have studied system fairness over time, these often focus on the distribution shift~\cite{koh2021wilds} arising from changing sub-populations or systems behavior~\cite{morik2020controlling}, whereas we focus on a setting where the population remains the same but data completion rates change over time. In the sections below, we describe both the policy background and the algorithmic fairness literature motivating this work\looseness=-1.

\subsection{Policy Background}
\textbf{Data Infrastructure for Measuring Health Care Disparities.}
 Landmark studies, such as the ``Heckler Report'' (1985) \citep{heckler1985report} and the Institute of Medicine's (IOM) ``Unequal Treatment: Confronting Racial and Ethnic Disparities in Health Care'' report \citep{institute_of_medicine_us_committee_on_understanding_and_eliminating_racial_and_ethnic_disparities_in_health_care_unequal_2003}, have centrally shaped our understanding of racial disparities in the U.S. healthcare system. Published nearly 20 years apart, these reports revealed the harmful impact of racial health disparities throughout the U.S., recommended improved data collection related to race, and advanced legal, regulatory, and policy interventions specific to the medical field. Following these reports, considerable research has focused on reducing health disparities \citep{baciu2017communities}.  In our work, we build on this research by advocating for the data infrastructure necessary for timely disparity assessments in primary care settings.\looseness=-1 

The current picture of health disparities --- at the national, state, or local level --- is limited by the quality of available demographic data ~\citep{bierman2002addressing}. Many challenges first identified by IOM persist, in part due to poor data collection~\citep{james2023using}. A recent \textit{Urban Institute} report from \citet{james2023using} details challenges to collecting data on race, including lack of trust from both patients and providers, limited community engagement, and fragmented and inconsistent data systems. Patients may be asked to provide their race multiple times, with varying standards for recording racial categories across institutions. Data collection efforts are frequently uncoordinated and siloed across different patient interaction points, such as hospitals and insurance plans. Often, collection of race information is simply not a priority, and there are few mechanisms in place to produce high quality demographic data collection. \textit{Overall, no widespread requirements for standardization and timely reporting of race in healthcare organizations exist. In particular, the goal of conducting regular disparity assessments for meaningful health outcomes, while acknowledged as worthwhile, has not been a consistent policy priority }\looseness=-1 \citep{national2004eliminating, james2023using}.

\textbf{Mandated Reporting at Federal and State Levels.}
Recent federally-led efforts to address such gaps have been slow and inconsistent~\citep{machledt2021addressing}. For example, since 2016, Medicaid has required states to develop health disparity assessment plans, which would require stratification by race. However, implementation by the Centers for Medicare and Medicaid Services (CMS) was itself delayed, and reporting eventually became voluntary \citep{watson2024community}. New regulations --- 42 C.F.R. § 437.10(b)(7), (d) (2023) --- will require states to report core health quality measures stratified by race starting in 2027 \citep{watson2024community}.
CMS also recently announced the \textit{Hospital Commitment to Health Equity} measure, which requires that hospitals participating in CMS programs report on whether they are prioritizing equity, but does not require systematic health disparity assessments~\citep{cms_framework}.\looseness=-1

More concrete advances have been led by states. Since 2011, Michigan has been reporting on health disparities annually, with programmatic efforts to reduce these disparities. Michigan Medicaid now links reimbursements and performance bonuses to reductions in health disparities across five measures: diabetes (hemoglobin HbA1c) testing, cervical cancer screening, child wellness visits, postpartum care, and chlamydia screening, an approach that aligns with \citet{james2023using}'s recommendations. In particular, they argue for tying accountability measures and incentives to the reporting of health disparities. California now requires the reporting of race information, although data collection is fragmented and lacks a universal standard. Legislation enacted in 2022 requires hospitals to prepare and submit annual health equity reports along with an action plan \citep{hemrp_ca}. The first reports will be due in mid-2025 and will involve annual reporting on health outcomes disaggregated by race, among other demographic characteristics \citep{hemrp_ca}. As numerous other states and localities propose similar initiatives to regularly report on disparities~\cite{manning2022massachusetts}, it will be increasingly important to consider, understand, and potentially mitigate the impact of data reporting delays on accurate assessments. Our study contributes a comprehensive framework towards these aims.\looseness=-1

\subsection{Machine Learning Background}
\textbf{Dynamic and Pipeline Aware Fairness.}
\label{sec:background_pipeline_fairness}
 Our work is also related to dynamic or longitudinal fairness~\cite{d2020fairness, ge2021towards} assessment where fairness of a sociotechnical system is assessed over time --  due to shifting populations~\cite{koh2021wilds} or system updates~\cite{morik2020controlling}. However, in contrast to prior work, we highlight that \emph{missingness} in critical data variables can occur dynamically due to delayed reporting (\textit{e.g.}, of race). 
 Another related literature is early stopping in clinical trials where preliminary measurements of health outcomes can lead to incorrect disparity assessments. Prior work~\cite{chien2022multi,adam2023should} has considered the fairness implications of early stopping and adaptivity in clinical trials.
 However, these settings differ from ours, in that it is assumed that demographic data is available for all patients upfront. 
 
 Our work underscores the importance of a \textit{pipeline-aware} machine learning~\citep{black2023toward} perspective. In the context we study, a \textit{pipeline-aware} perspective entails systematically interrogating data collection and reporting systems, as well as the way missing or damaged data is handled in data preprocessing, and how imputation might be performed. We identify delay as a potential blindspot in machine learning pipelines, as it is only recognizable across longitudinal snapshots of individuals, while the bulk of the literature has analyzed static datasets. Our work shows the need for explicitly examining reporting delays in decision-making pipelines.

 \looseness=-1

\textbf{Missingness in Demographic Information.}
Missing data and imputation are well-studied topics in statistics and sociology \citep{little2019statistical, little1989analysis}. The impact of missing data, particularly race data, has been widely studied in health research as well \citep{dembosky2019indirect, derose2013race, bierman2002addressing, spangler2023missing, branham2022trends, polubriaginof2019challenges, fremont2016race}. Within the algorithmic fairness literature, there has been considerable attention paid to the consequences of missing features -- including sensitive information \citep{mitchell2021algorithmic, zhang2021assessing,akpinar2024impact, khan2024still, awasthi2021evaluating} and missing data imputation \citep{jeanselme2022imputation, fernando2021missing, zou2023implications}. \citet{zhang2021assessing} provide theoretical bounds on fairness estimation error in the presence of missing data. \citet{fernando2021missing} document several missingness patterns such as item non-response and attrition, and find that imputing rows with missing data can mitigate bias. However, they do not consider \emph{delay} in their discussion of missing data.\looseness=-1

\citet{jeanselme2022imputation} study multiple forms of missingness processes and emphasize that no single imputation strategy outperforms across all processes. While they do not characterize delay, the implication of their findings is that delay, when it exists, would also manifest in its own unique patterns, further complicating efforts to address missingness through imputation. \citet{akpinar2024impact} study the systematic problem of ``differential feature under-reporting'': a phenomenon in which some data records are more likely to be complete for individuals who interact with the system more frequently. They show that under-reporting tends to exacerbate disparities and propose mitigation methods. Our work assesses whether delay is also differential in similar or dissimilar ways across patterns of care-seeking behavior.

Our focus specifically on delays provides a novel opportunity to advance the missing data literature. Not only are delays on their own an important source of missing data to consider in real-world applications, by definition, they produce data that is only missing for some period of time. In other words, the ability to validate the ground truth of delayed data might shed light on some of the mechanisms that contribute to missingness not at random (MNAR), which otherwise are not observable to researchers within a static dataset \citep{jeanselme2022imputation, awasthi2021evaluating, khan2024still}.\looseness=-1

\section{Data}
\label{sec:data}
We leverage access to the American Family Cohort (AFC) dataset, which contains data from over 1,000 practices, all of which are part of the American Board of Family Medicine (ABFM) PRIME Registry \citep{abfm_data}. The PRIME registry functions as an intermediary between healthcare providers and the Centers for Medicare \& Medicaid Services (CMS). They help with collecting and analyzing data, and produce quality measures on behalf of clinicians for incentive-based programs managed by CMS \citep{cms_measures}. In contrast to many conventional machine learning datasets, these data contain fine-grained longitudinal information of patient interactions (see Figure~\ref{fig:cohort-schematic}), including changes in race reporting, which enables us to conduct realistic assessments of the magnitude, correlates, and impact of reporting delays.

\textbf{Data Collection and Incentives for Disparity Assessments.} Healthcare practices that join the PRIME registry have access to a detailed set of dashboards with information about their practice service area, disease prevalence, and care quality gaps. 
Many practices share data with the registry because they do not have the capacity to do their own analyses and reporting to be in compliance with CMS. 
As previously noted, while CMS does not yet require racial health disparities to be analyzed and reported, local programs and mandates are beginning to emerge.
ABFM and partnering researchers are well-positioned to conduct disparity assessments using health data on behalf of practices in the registry.
Our study is a practical demonstration of this, specifically the potential impacts of delayed reporting.\looseness=-1

\textbf{Key Features of Dataset.} The AFC dataset is ideally suited for studying reporting delays.
First, this data contains longitudinal information including patient demographics, visits, diagnoses, observations, and procedures, as well as some clinician-specific details. Second, practices from all 50 states are represented in the data. Third, the data includes significant representation from healthcare practices and patients with both private and public insurance plans, as well as distinct electronic health records (EHR) systems. 
These characteristics make AFC data a meaningful and realistic test case for studying racial disparities across the U.S., as opposed to analyses that may focus on less diverse sub-regions, specific providers, single EHR systems, or a small subset of medical conditions. Summary statistics for this dataset are in Table~\ref{table:app-balance-table}.\looseness=-1

Most importantly for our study, the data is \emph{longitudinal} and information updates are timestamped, providing the possibility of observing the phenomenon of delay that would otherwise be hidden. Every time information is modified or added for a patient, a new record is added to the AFC dataset with a timestamp and linkable patient ID (see Longitudinal Reporting in Figure~\ref{fig:cohort-schematic}), without over-writing previous timestamped records for the same patient. 
This includes cases when demographic data such as race is updated. As a result, we can track the reporting of race for each patient over time and produce estimates of delayed race reporting, differentiating this dataset from other datasets with a static availability of race per patient. These timestamps come from the data provider, and the cadence of the updates does not always follow a regular pattern. In particular, some practices \textit{push} their data -- meaning they submit the data to the registry -- while others experience data \textit{pulls} at regular intervals.

For large scale audits, by the time patient information is aggregated into the AFC dataset, \textit{any upstream source of delay} in reporting of race --- whether due to patient hesitance, failure to request the information at the time of the patient visit, or data collection lags on the part of the data provider --- creates delay that can materially affect the quality of disparity assessments. Therefore, we focus on the \textit{consequences} of delays rather than the precise \textit{causes} of the delays. We further define reporting delays in the next section.

\section{Methods}
\label{sec:methods}
In this section, we describe how we define and quantify the impact of delays on disparity assessments\footnote{Code: https://github.com/reglab/delayed-reporting}.
\subsection{Defining Delays}
\label{sec:defining-delay}
We define \emph{race reporting delays} using a time-based measure of delay. Delayed reporting occurs if there is a gap between the earliest possible date of reporting, and when race is actually reported. We operationalize the former as whichever occurs latest among (1) the earliest \new{timestamp denoting when the} patient’s date of birth is reported or (2) the earliest per-practice race reporting (\textit{i.e.}, the first patient in a practice with race information). 
We consider a patient's race to be \emph{reported} if there is a non-missing race or ethnicity entry that corresponds to one of the federal race and ethnicity categories (as described in Section~\ref{sec:data-processing}). Importantly, our date information reflects the date when this information was shared or updated with the data provider responsible for producing the AFC data, not the clinical encounter when the patient may have self-reported race. As a result, this date may lag in comparison to the true clinical encounter. 
However, this definition of delay still captures realistic data lags (\textit{e.g.}, due to a range of behavioral, administrative, and technological factors) that an independent evaluator would encounter when conducting a disparity assessment.

\subsection{Data Processing}
\label{sec:data-processing}

\subsubsection{Measurement of Race and Ethnicity}
\label{sec:measurement-race}
We parse and code categorical versions of patient race and ethnicity from both free-text and categorical race-related fields in the AFC data, following the same processing steps as Cheng et al.~\cite{cheng2023redundant}.
To harmonize across a wide range of data schemas, we map all entries to the $1997$ Office of Management and Budget (OMB) federal standard for race and ethnicity reporting: American Indian or Alaska Native (AIAN), Asian, Black, Native Hawaiian or Pacific Islander (NHPI), White, Multiracial, Other, and Hispanic \citep{omb1997}.\footnote{Note that the race group of ``Middle Eastern or North African'' was only added in the 2024 OMB categories update~\cite{omb_update}.}
Following OMB standards, we record patients as Hispanic or Latino if they indicate their ethnicity as such, in addition to their indicated race. For analyses involving prevalence estimates, we combine Asian and NHPI into the Asian and Pacific Islander (API) category. To detect when race is unknown or declined, we use string matching to a curated set of keywords to identify data points with no reported race (see  Appendix~\ref{sec:unknown_race_parsing}). For patients who report race multiple times ($<1\%$), we parse their first reported race --- matching the time at which we consider race to be reported. \new{Note that we expect detection rates for multi-racial patients to be lower than those for other racial groups, as only simple regex parsing rules are applied.} \looseness=-1

\subsubsection{Cohort Definition}
From the full AFC dataset of 7.8M patients, we restrict analysis to patients for whom we ever have a recorded race mappable to OMB categories, and we identify the earliest date at which that recorded race is available. %
 Because our objective is to assess the prevalence and impact of \textit{delays} (\textit{i.e.}, cases for which we can eventually recover a race recording), we also exclude patients who \textit{never} report race, or whose race cannot be parsed using our automatic processing techniques ($\sim$900k patients). We only consider patients who are $\geq$18 years in 2018, which is the primary year we use for most analyses. Our final cohort consists of $5,310,700$ adult patients whose race is recorded with either some or no delay (as of early 2024). We then identify whether a patient has experienced a reporting delay by producing a continuous measure reflecting the number of days from a patient's earliest reporting opportunity up until the date that race is in fact available (Section~\ref{sec:defining-delay}).

\subsubsection{Health Outcomes and Metadata}
\label{sec:health_outcomes}

\label{feature-extraction}
We observe patient attributes that have been standardized and cleaned according to the Observational Medical Outcomes Partnership (OMOP) Common Data Model.\footnote{\url{https://www.ohdsi.org/data-standardization/}} In addition to race (as described in Section~\ref{sec:measurement-race}), we extract patient age, sex, and marital status.
We also extract clinical information such as the number of patient visits, the length of time they have interacted with a practice, and health-related outcomes like disease diagnoses, procedures, and observations. Like demographic characteristics, all of these features follow the OMOP data model. Like race information, some information is subject to reporting delays and missingness. For understanding the AFC population (see Table~\ref{table:app-balance-table}), we treat these attributes as fixed (\textit{i.e.}, we extract attribute information if it ever appears in the AFC data) and do not consider the impact of delays beyond delays in race reporting.

In the context of disparity assessments, we compute six binary health outcomes: three condition diagnoses (depression, diabetes, and hypertension), two procedures (electrocardiograms and depression screens), and a clinical observation (hemoglobin HbA1c tests). We curate these health outcomes based on prior literature which provides evidence of racial disparities (see more details in Appendix~\ref{sec:extract_health_outcomes}).

\subsection{Retrospective Analysis on the Impact of Delays}
\label{sec:estimatating_impact_metrics}
Drawing on the definition of delay in Section~\ref{sec:defining-delay} and the health outcomes described in Section~\ref{feature-extraction}, we next examine the impact of reporting delays via a retrospective analysis of disparities. We consider disparity assessments to include any comparison of health outcome prevalence by racial group, for a particular time period (\textit{e.g.}, the White-Black hypertension diagnosis gap in the first quarter of 2018). %
Since we have timestamps of when race information became available for each individual patient, we can demonstrate what a particular disparity assessment would have looked like \textit{retrospectively}, if it had been conducted at any previous time point. For instance, we can simulate a disparity assessment for 2018 Q1 immediately following its conclusion, at which point 40.94\% of patients have delayed race information and are thus excluded from prevalence calculations by racial group. The same disparity assessment for 2018 Q1, conducted with the benefit of more hindsight (\textit{i.e.}, using more complete race data provided after delay, but health outcomes remaining fixed for 2018 Q1), could yield different results because more patients would be included in the analysis given their race availability. Our core objective is to isolate this impact of reporting delays on health disparities.

We conduct these analyses at three distinct geographic levels: national, state, and practice-level. First, with simulations at the national level, our goal is to understand how delayed reporting of race may affect aggregated disparity estimates similar to annual reports like the ``National Healthcare
Quality and Disparities
Report''~\citep{ahrq}. Second, as noted above, specific states, such as California and Michigan, have been pushing for deeper assessments of health disparities, so we also conduct analyses at the state level. Simulating disparity assessments may yield more variation at the state level, particularly as some states may experience more delays than others. We may also more easily observe distinct trends in delayed reporting that are overridden at the national level. Lastly, because administration of race reporting and mitigation efforts occur within physician practices, we also study the impact of reporting delays at the practice level.

\subsection{Metrics for Error in Disparity Assessments}
\label{sec:metrics}
In all cases, we measure prevalence, or rates of occurrence, of health outcomes. We introduce two primary error metrics of interest based on changing accuracy of health monitoring, as race information becomes more complete over time: \emph{prevalence errors} and \emph{disparity errors}. Prevalence errors are the differences in prevalence estimates for racial groups in a cohort (defined by a fixed time period such as 2018 Q1) at some initial time point $t_\text{initial}$ (the first possible assessment of the cohort, when race information is most incomplete) compared to time point $t_{\text{final}}$ (when all race information is known for the cohort).\footnote{\new{We also report relative absolute prevalence error in Section~\ref{sec:relative-absolute-error}, which is a common metric used to evaluate prevalence under class imbalance~\citep{esuli2023learning}.}} We also visually present prevalence estimates at quarterly intervals past $t_\text{initial}$ to show how error is reduced over time as more race information is collected (see Figures~\ref{fig:simulation} and~\ref{fig:simulation-state}). Disparities are the pairwise comparisons of prevalences between two racial groups, and so disparity errors derive from, but do not necessarily appear the same as, prevalence errors (\textit{i.e.}, prevalence errors in the same direction may yield no disparity error). 
We summarize all metrics in Table~\ref{tab:metrics_equations}. To obtain uncertainty estimates, we bootstrap by resampling $50$ times across both practices and individual patients, and averaging metrics across all bootstrapped samples.

\begin{table*}
\resizebox{\textwidth}{!}{%
\centering
\begin{tabular}{l|c|c}
\toprule
Metric & Definition & Equation\\
\toprule

$\textbf{prevalence}_{(j,t)}$ & Rate at which outcome $Y$ occurs, estimated for any group $j$ at any time $t$ & $
 \frac{\sum_{i \in j} Y_i}{\sum_{i \in j} 1}
$\\
$\textbf{weighted prevalence}_{(j,t)}$ & Prevalence weighted by posterior probability $p_{ij}$ when race is unobserved  & 
$\frac{\sum^{N}_{i}p_{ij} \cdot Y_i}{\sum^{N}_{i}p_{ij}}$\\

$\textbf{prevalence error}_j$ & Difference between initial and final prevalence for group $j$ & $\text{prevalence}_{(j,{t_{\text{final}})}} -  \text{prevalence}_{(j,{t_{\text{initial}})}}$\\

$\textbf{relative absolute prevalence error}_j$ & Absolute difference between initial and final prevalence for group $j$ relative to final prevalence & $\frac{|\text{prevalence}_{(j,{t_{\text{final}})}} - \text{prevalence}_{(j,{t_{\text{initial}})}}|}{\text{prevalence}_{(j,{t_{\text{final}})}}}$\\

\textbf{average prevalence error} & Absolute prevalence error averaged across all groups & $\frac{1}{G}\sum_j^G| \text{
prevalence error}_j|$\\

$\textbf{disparity}_{(j,j+1,t)}$ &  Difference in prevalence between groups $j$ and $j+1$ at any time $t$ & $\text{prevalence}_{(j,t)} - \text{prevalence}_{(j+1,t)}$\\

$\textbf{disparity error}_{(j,j+1)}$ & Difference between initial and final disparity for two groups & $\text{disparity}_{(j,j+1,{t_{\text{final}})}} - \text{disparity}_{(j,j+1,{t_{\text{initial}})}}$\\

$\textbf{average disparity error}$ & Absolute disparity error averaged across all pairwise group combinations & $\frac{2}{G(G-1)}\sum_j^G\sum_{j+1}^G |\text{disparity error}_{(j,j+1)}|$ \\

\bottomrule
\end{tabular}}

\caption{Error metrics for prevalence and disparity assessments from time $t_\text{initial}$ (maximum number of patients with delayed reporting) to $t_\text{final}$ (race fully known). $Y$ denotes a health outcome, with $Y_i \in \{0,1\}$  indicating the presence or absence of the outcome for an individual $i$ at time $t$, in a population of $N$ patients. $j$ denotes an individual racial group (and $j+1$ a different racial group), up to $G$ total racial groups. $p_{ij}$ denotes the posterior probability of an individual $i$ (where $0\leq p_{ij} \leq 1$) belonging to a specific racial group $j$. \label{tab:metrics_equations}}
\end{table*}

\section{Results}
\label{sec:delayed-reporting-desc}

We begin by reporting results on the prevalence and correlates of delays. We then report results from our retrospective analysis described in Section~\ref{sec:estimatating_impact_metrics} and calculate the error metrics described in Section~\ref{sec:metrics}. We find that the impact of reporting delays is substantial. Because of the richness of the dataset, we distill core results here, and provide more detailed results in the Appendix.\looseness=-1

\textbf{Delays are the norm, not the exception.} Over 73\% of patients ($N=3,911,213$) in our cohort experience some delay, and over half experience delays $\geq$ 60 days. Overall, 21 states and over half of practices exhibit a similar degree of delay (75\% of patients or more), indicating that the phenomenon is both widespread and consistent. Put differently, any well-intentioned efforts to conduct routine, quarterly assessments (\textit{i.e.}, within three months of the health outcomes in question) would likely discard a majority of all patients from analysis. 

\begin{figure}[htbp]
\centering
\includegraphics[width=\linewidth]{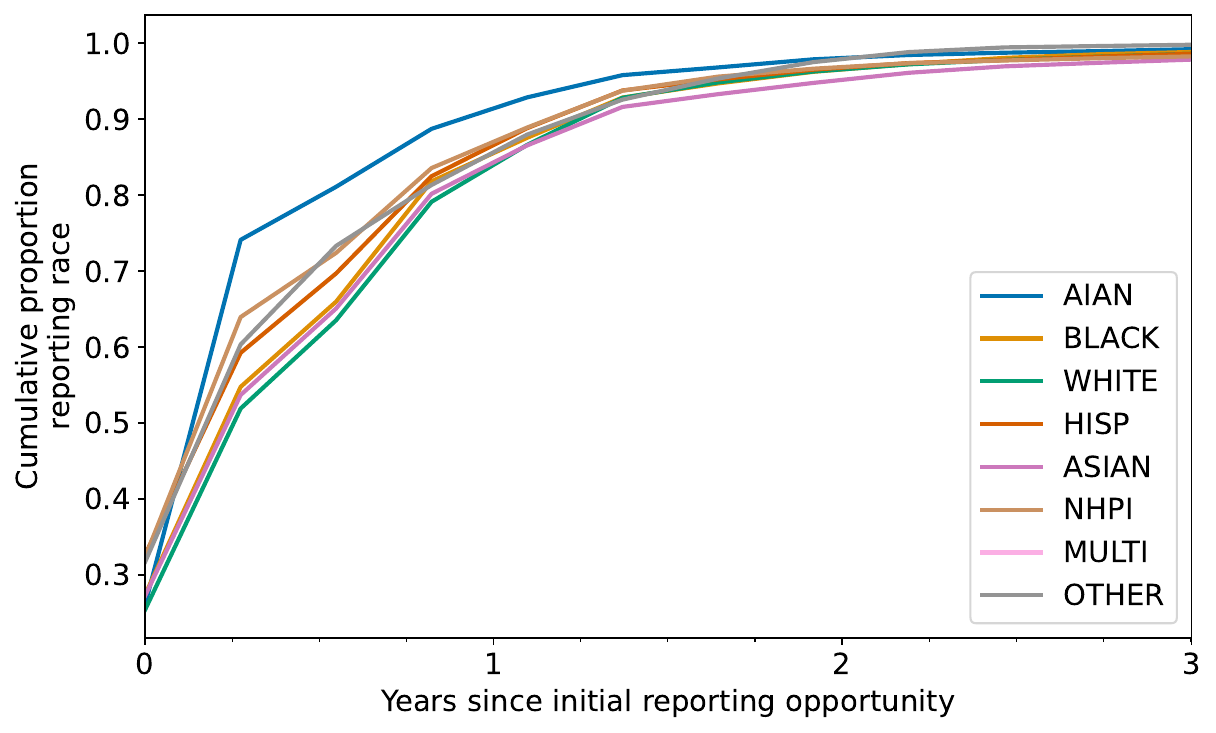}
\caption{\textbf{Reporting rates differ by race and ethnicity.} On the x-axis, we show the time difference from the earliest date at which a patient could have reported race and ethnicity up until the date at which their race and ethnicity is \emph{known}. To see the differences more clearly, we only depict the first three years on the x-axis. The cumulative proportion of patients within each racial and ethnic group who have reported race and ethnicity is on the y-axis.}
\Description{Lines for AIAN and NHPI groups appear steeper initially, indicating higher rates of reporting, compared to other race groups.}
\label{fig:cumulative-dists}
\end{figure}

\textbf{Delayed reporting of race does not affect all groups evenly.} If patients with timely reported data are representative of all patients, reporting delays may not pose a substantive problem. But, delays do not affect groups equally. Figure~\ref{fig:cumulative-dists} shows that the cumulative rates of reporting are much steeper for racial groups like AIAN and NHPI, while White, Black, and Asian groups appear to experience greater lags. Kruskal-Wallis tests for all pairwise comparisons are statistically significant with $p<0.01$, even with Benjamini-Hochberg adjustment for multiple comparisons. In short, reporting delays are not only pervasive, but themselves have a distributive dimension across race.

\begin{table}[htbp]
\raggedleft
\centering
\begin{tabular}{p{0.6in}S|SS}
\toprule
 &  \textbf{Overall Average} &  \textbf{No Delay} &  \textbf{Delay}    \\
\midrule 
    N & \text{5,310,700} & \text{1,399,487} & \text{3,911,213} \\
\midrule

Age (years) & 58.02 & 55.97 & 58.76 \\
Female (\%) & 56.26 & 55.34 & 56.59\\
Male (\%) & 43.69 & 44.58 & 43.37 \\
Other (\%) & 4.69 & 7.61 & 3.65  \\
AIAN (\%) & 0.73 & 0.73 & 0.74 \\
Asian (\%) & 2.69 & 2.78 & 2.65 \\
Black (\%) & 8.45 & 8.78 & 8.33 \\
NHPI (\%) & 0.52 & 0.66 & 0.47 \\
White (\%) & 80.25 & 77.49 & 81.24 \\
Hisp (\%) & 10.91 & 13.52 & 9.98 \\
Other (\%) & 4.12 & 4.78 & 3.89 \\
Multi (\%) & 0.89 & 0.93 & 0.87 \\

\bottomrule
\end{tabular}

\captionof{table}{\textbf{On average, patients who experience delays are older and more likely to be White.} We show differences in average demographic characteristics between patients who experience delays compared to patients with no delays. All differences are statistically significant ($p<0.01$) with multiple testing corrections except AIAN (\%). Also see Appendix~\ref{table:app-balance-table}.}\label{tab:ds_summary_stats}
\end{table}

\textbf{Patients with delayed race reporting are older, more care-seeking, and less healthy.} We also find that reporting delays are correlated with a wide range of other patient attributes, threatening the validity of static disparity assessments. Tables~\ref{tab:ds_summary_stats} and~\ref{table:app-balance-table} present differences in means across a variety of patient-level characteristics, almost all of which are statistically significant. White patients are over-represented among those with delays while Hispanic patients are under-represented.
Patients with delays also have more visits on average, and their first visit occurs earlier in the data. Lastly, patients with delays tend to be less healthy, with higher rates of all six health outcomes we measure. For example, they are more than 10 percentage points more likely to receive a diabetes (HbA1c) test.

\textbf{Delay is also associated with differences in practice-level characteristics related to data collection and data management.} Patients without delays are more likely to come from practices using Practice Management (PM) systems. PM systems automate many billing and administrative tasks, which can include data collection of demographic information \citep{pm_systems}. This finding aligns with prior work that suggests integrating EHR and PM systems may lead to improved data collection \citep{carter2008electronic}. Delays are also associated with practices that have their data only ``pulled'' --- meaning the registry initiates extracting data on some regular cadence. 
Patients without delays are more likely to come from practices that combine data pulls with data ``pushes'' and other update methods --- likely closer to real-time updates. The relationship between delays and specific EHR systems is particularly strong, while having multiple EHR systems is more likely among patients with no delay (Table~\ref{table:app-balance-table}).

\begin{figure*}
    \centering
    \includegraphics[width=0.8\linewidth]{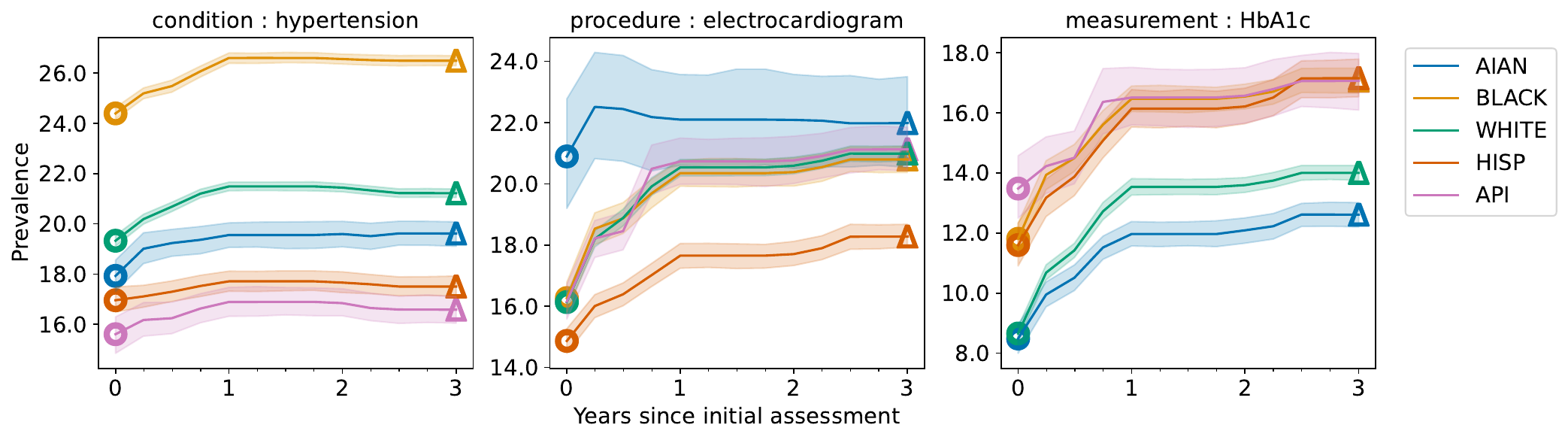}
    \caption{Simulations at the national level for one condition (hypertension), one procedure (electrocardiograms), and one measurement (HbA1c for diabetes). See additional outcomes in Figure~\ref{fig:simulation-appendix}. 
    If delayed reporting had no effect on prevalence, we would expect to see horizontal lines for each race line within facets. Instead, each facet shows that rank orderings of prevalence by race changes over time, and prevalence by race often increases monotonically. Additionally, there is high uncertainty for some estimates such as electrocardiogram procedures, and those estimates experience the most fluctuation over time. All prevalence estimates are conducted on a fixed cohort of patients from 2018 Q1.}
    \Description{Prevalence estimates at the national level change over time due to delays in race reporting. On the x-axis, the number of years from the initial assessment date (2018 Q1) is shown. Prevalence estimates for each health outcome disaggregated by race  appear on the y-axis. 95\% confidence intervals represent uncertainty. In general, prevalence estimates for most race groups increase over time as more race information is recovered.}
    \label{fig:simulation}
\end{figure*}

\textbf{Delayed reporting \new{can obfuscate measurement of prevalence.}}
\label{errors}
Even if reporting delays are pervasive and unevenly distributed, do they affect the estimation of health outcomes? 
We find that using prevalence estimates that do not account for delays can significantly \emph{distort} time-sensitive monitoring of health-related outcomes. We illustrate this phenomenon by calculating prevalence rates for a cohort of $1,776,729$ patients from 2018 Q1. This is a subset of all $\sim5.3$M patients in our study, as we restrict to (1) patients whose date-of-birth is reported before 2018 (ensuring minimum patient data robustness for the time period in question) and (2) practices that report race before 2018 (eliminating practices that were not collecting race at all). We can produce estimates first at $t_\text{initial}$ immediately following the conclusion of the quarter, when only 59.06\% of patients have recorded race, then at regular intervals up to $t_{\text{final}}$, when 100\% of patients have recorded race. Since our disparity assessment remains exclusively focused on 2018 Q1, health outcomes in the cohort are held fixed based on those occurring in that quarter; the only change across iterations is the proportion of patients omitted due to missing race information (which decreases over time). Table~\ref{tab:prevalence-rates} summarizes changes in prevalence and disparity estimates, attributable entirely to delayed race information, between $t_{\text{final}}$ and $t_{\text{initial}}$ (where a number closer to zero indicates a lower error). For example, diabetes (HbA1c) testing prevalence error among Hispanic patients is $5.56$ percentage points, which is \new{higher than for} other racial groups. This means that the true prevalence estimate, measured at $t_\text{final}$, is around $5$ percentage points higher than the prevalence estimate at $t_\text{initial}$.
Most error values are positive, consistent with our finding from Table~\ref{table:app-balance-table} that greater delay is associated with higher prevalence of health outcomes. 
The average prevalence error across all groups and all outcomes is 2.15 percentage points. Figure~\ref{fig:simulation} visualizes these trends at quarterly intervals from right after 2018 Q1 up to 3 years later ($t_{3Yrs}$), when 99.85\% of patients have recorded race and the average prevalence error has been reduced to less than 0.1 percentage points. The effects of delays across multiple consecutive cohorts (2018 Q1, Q2, Q3, and Q4) are detailed in Appendix~\ref{sec:continuous-monitoring}\looseness=-1.

\begin{figure*}
    \centering
    \includegraphics[width=0.8\linewidth]{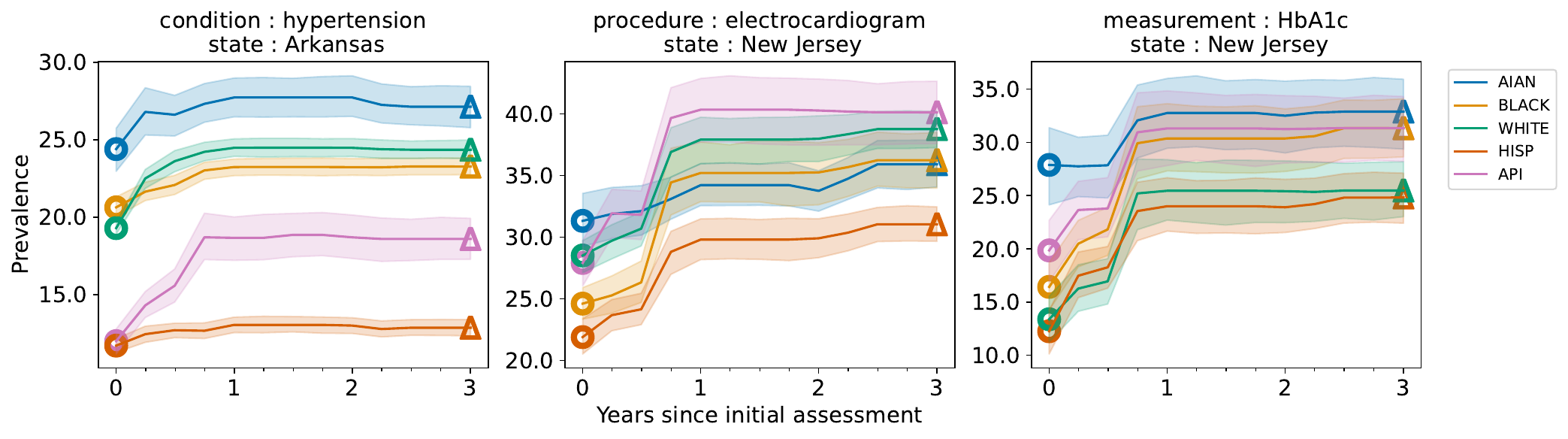}
    \caption{Simulations at the state level for hypertension diagnoses, electrocardiogram procedures, and HbA1c tests. Changes in prevalence estimates over time differ in magnitude across races, indicating variability in disparities across race groups. The states shown are each chosen from among the top three states with the highest average number of patients for each health outcome. See additional state-level outcomes in Figure~\ref{fig:simulation-state-appendix}. All prevalence estimates are conducted on a fixed cohort of patients from 2018 Q1.}
    \Description{Prevalence estimates at the state level (for Arkansas and New Jersey) change over time due to delays in race reporting. On the x-axis, the number of years from the initial assessment date (2018 Q1) is shown. Prevalence estimates for each health outcome disaggregated by race appear on the y-axis. 95\% confidence intervals represent uncertainty. In general, prevalence estimates for most race groups increase over time as more race information is recovered.}
    \label{fig:simulation-state}
    \vspace{-0.5em}

\end{figure*}

We observe similar discrepancies at the state and practice level. 37 states have the same or higher amount of average prevalence error as seen at the national level, while California in particular has a lower average prevalence error (1.20 percentage points). Figure~\ref{fig:simulation-state} illustrates state-level examples of the 2018 Q1 assessment and underscores the heterogeneity and unpredictability of delayed reporting's effects across different geographies. As shown in Figure~\ref{fig:practice-dist-differences} in the Appendix, most prevalence estimate errors at the practice level are small and clustered around $0$, though there are a non-negligible number of outliers.
Across over 1,000 individual practices, 13.39\% of all prevalence estimates for each race group and health outcome are incorrect by over 10 percentage points. %

\textbf{Delayed reporting distorts true disparity in retrospective analyses.}
\label{simulations}
Do these prevalence errors affect estimates of \emph{disparities} between demographic groups?  As seen in Figures~\ref{fig:simulation} and \ref{fig:simulation-state}, even small absolute differences can lead to changes in the relative magnitude of group-level disparities. For example, an early assessment of Arkansas hypertension prevalence in 2018 Q1, conducted at $t_\text{initial}$, would lead one to conclude there is virtually no disparity between API and Hispanic patients. However, with more complete race information available after three years, one can see that the API-Hispanic hypertension gap was actually more than 5 percentage points in that quarter, with no overlap in 95\% confidence intervals. 
Similar minimizations of disparities occur at the national level for the Hispanic-AIAN hypertension gap, the White-AIAN diabetes (HbA1c) testing gap, and the Hispanic-Black diabetes gap (see Figure~\ref{fig:simulation-appendix}), among others.
As further detailed in Table~\ref{tab:prevalence-rates} and Appendix~\ref{app:ex_min_sign}, the magnitude and direction of disparity error 
varies across pairwise comparisons, across outcomes, and across geographic scales of assessment.

\begin{table*}
\centering
\begin{tabular}{lSSSSSSS}
   {\textbf{Health outcome}} & \multicolumn{5}{c}{\textbf{Prevalence error}} & {\textbf{Average prevalence error}} & {\textbf{Average disparity error}}  \\
   \cline{2-6}
 & \text{AIAN} & \text{API} & \text{Black} & \text{Hispanic} & \text{White} & & \\
  \hline
Diabetes & 0.88 & 0.86 & 1.12 & 0.26 & 0.75 & 0.81 & 0.57\\
Hypertension & 1.64 & 0.96 & 2.09 & 0.54 & 1.88 & 1.49 & 1.05\\
Depression & 1.63 & 0.35 & 0.71 & 0.77 & 1.18 & 0.93 & 0.64\\
Depression screen & 0.97 & 1.25 & 0.87 & -0.18 & 1.38 & 1.02 & 0.86 \\
Electrocardiogram & 1.07 & 5.00 & 4.53 & 3.41 & 4.85 & 3.86 & 2.06\\
HbA1c & 4.14 & 3.60 & 5.26 & 5.56 & 5.34 & 4.78 & 1.34 \\
\hline
\end{tabular}
\caption{Errors in prevalence and disparity estimation between $t_\text{initial}$ and $t_\text{final}$ at the national level for each racial group and health outcome, focused on health outcomes for a cohort in 2018 Q1. Values provided are percentage points; \textit{i.e.}, premature assessment of Hispanic HbA1c tests underestimates prevalence by 5.56 percentage points. See Table~\ref{tab:metrics_equations} for definitions of metrics and Table~\ref{tab:relative-absolute-error} for relative absolute prevalence error. Most prevalence errors are statistically significant ($p< 0.05$) with multiple testing correction.}
\label{tab:prevalence-rates}
\vspace{-1em}
\end{table*}

\textbf{Imputation is not a panacea for delayed reporting.} One approach to mitigate the effect of delayed race information might be to \emph{impute} the posterior probability of a patient's race, as is a common strategy for missing data in general. We explore whether widely used imputation methods such as Bayesian Improved First Name Surname Geocoding (BIFSG) can improve the accuracy of disparity assessments performed prior to obtaining complete race information for all patients \citep{elliott2009using, imai2016improving, voicu2018using}. Bayesian methods that predict individuals' race using a combination of first names, last names, and geography have been used across various domains to evaluate disparities \citep{elliott2009using, zhang2018assessing, hepburn2020racial, elzayn2024measuring}. Following conventional practice \citep{elzayn2024measuring, chen_fairness_2019}, in order to calculate a BIFSG version of prevalence, we weight each patient's contribution to each racial group's prevalence by their posterior probability $p_{ij}$ of being in that group (see Section \ref{sec:metrics}). 

\begin{figure*}[htb!]
\centering
\includegraphics[width=0.95\linewidth]{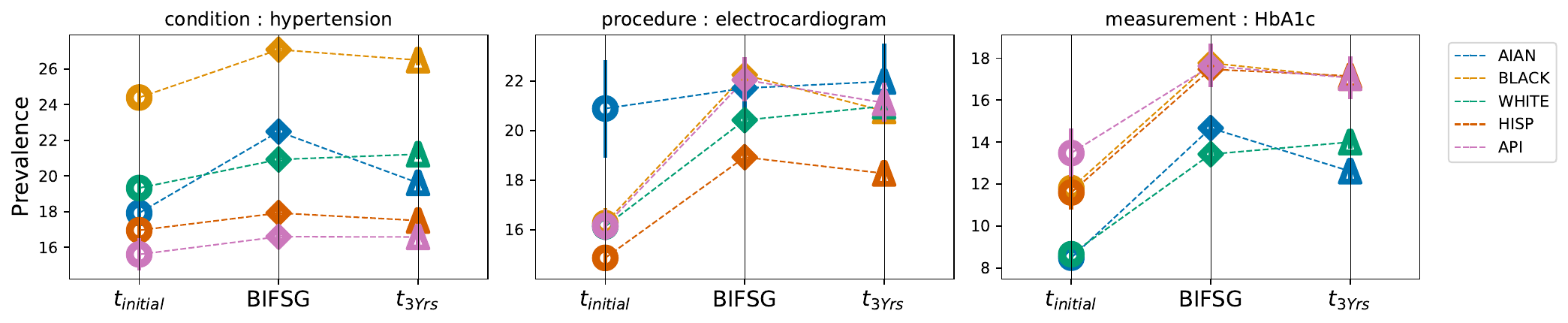}
\caption{BIFSG over-estimates prevalence for several minority race groups (\textit{e.g.}, Black and AIAN patients) across several outcomes, though average prevalence error is improved. For example, prevalence estimates with BIFSG for Black patients (yellow diamonds) are higher than estimates at $t_{initial}$ (yellow circles) and $t_{3Yrs}$ (yellow triangles), but are closer to estimates at $t_{3Yrs}$ (yellow triangles). In each subplot, the y-axis denotes the estimated prevalence. Values for $t_{initial}$ and $t_{3Yrs}$ match the same national values as shown in Figure~\ref{fig:simulation}. See additional outcomes in Figure~\ref{fig:bifsg_impute_all}.  
}

\label{fig:bifsg_impute}
\end{figure*}

At the individual level, BIFSG achieves AUROC values $>65\%$ for all racial groups. However, performance varies substantially across groups, with AUROCs ranging from 93.6 for Hispanic patients and 67.5 for AIAN patients,\footnote{We note that BIFSG AUROC for patients in `Other' race groups was low (34.22). For fair comparison with BIFSG, we exclude this group from all simulations.} suggesting group-level estimates may vary in accuracy~\cite{chernenko2023limits}.  %
We can see this problem in Figure~\ref{fig:bifsg_impute}, where prevalence estimates based on imputed probabilities from BIFSG are often over-estimated for minority groups (prevalence error is negative) though never for the majority White group. 
As a result of such over-estimation,  BIFSG does not consistently reduce delay-based errors in disparities, \new{which aligns with prior work on missing race or gender data~\citep{chen_fairness_2019}}. We perform one-sided Mann Whitney U-tests comparing error metrics between estimates of prevalence using BIFSG and $t_{3Yrs}$ versus $t_{initial}$ and $t_{3Yrs}$, with Benjamini-Hochberg correction for multiple testing. We observe that the average disparity error (detailed in Section~\ref{sec:estimatating_impact_metrics}) is only significantly improved (at 0.05 level) for diabetes diagnoses, electrocardiogram procedures, and HbA1c measurements. However, the average prevalence error does significantly decrease with the use of BIFSG for all outcomes, though the size of the difference is numerically small for some outcomes (see Figure~\ref{fig:appendix_bifsg_impute} in the Appendix). Note that results are sensitive to rounding of $p_{ij}$. These results indicate that imputation methods like BIFSG can mitigate delayed reporting to some degree with regard to prevalence errors, but do not produce accurate prevalence point estimates nor disparity error estimates. Thus, BIFSG cannot always replace accurate, self-reported race information in prevalence and disparity estimation.\looseness=-1

\section{Discussion: Implications for Practitioners}
\label{sec:discussion}
Our findings inform how to improve the use of data-driven decision-making tools in light of demographic reporting delays. 

\textbf{More holistic efforts should be made to understand and address the mechanisms driving delays.}
While there is an extensive literature on accounting for and imputing missing data in healthcare, the impact of changing missingness over time is less studied. Thus, modeling missingness mechanisms is an interesting direction of future work. In our case, this might involve efforts to uncover why delayed reporting is correlated with health outcomes, and the precise mechanisms through which delays occur during the patient intake and reporting process. This recommendation touches on a key limitation of our study, which is a retrospective analysis of a de-identified, pre-existing dataset. Given that we do not have the ability to contact the individual decision-makers  (\textit{e.g.}, nurses, clinicians, intake coordinators, etc.), we are unable to explore the many upstream factors that may have caused delayed race reporting. 
More qualitative analyses are essential to uncover the drivers of delayed reporting. For example, surveys could be conducted across practice sites to understand common data collection protocol and infrastructural reasons for delays.

One implication of our findings is that different EHR systems vary in their patient delayed reporting rates, suggesting that 
user design choices, as well as backend software architectures, may contribute to delays --- a meaningful overlap between human computer interaction (HCI) and fairness domains. 
Since rates of delay also vary across racial groups, these efforts should further consider the role of individual behavior --- such as hesitance to report race --- and practice-side variation in recordkeeping. We build on calls in the social sciences for incorporating qualitative research methods in data science work \citep{grigoropoulou2022data}, and complement existing efforts to document and understand practitioner experiences with data collection \citep{andrus2021we}, with appropriate data protection mechanisms (and communication to patients thereof) \citep{king2023privacy}. This recommendation also aligns with the growing recognition in algorithmic fairness that decision-making tools should be studied in their institutional contexts \citep{veale2018fairness, yang2016investigating}.\looseness=-1

\textbf{Understanding delays retrospectively is complex and challenging to test.}
Although we cannot study the precise mechanisms of delays, we hypothesize several mechanisms through which delays might occur, and why there are higher rates of delay among White patients and patients with more health conditions.

\textit{Patient hesitance:} While we cannot directly measure hesitance, we study patients whose reported race changes from ``declined'' or ``unknown'' to ``known.''  \new{We find only small differences in the proportion of White patients in this group relative to other patients in the data, which suggests that hesitance does not play a significant role in the delays that we observe here.}

\textit{Systemic Complexity and Delayed Presentation:} Patients with complex health conditions might have more complicated medical records and/or more administrative tasks \citep{kyle2021patient}, thus potentially leading to administrative delays in fully completing demographic information \citep{natlassoc}.

\textit{Intake / registration visits:} Some patient visits might be intake visits, where race data may be missing when these are not properly administered. Prior research suggests that Black and Hispanic patients are more likely to utilize the emergency department \citep{hong2007emergency, arnett2016race} and may face greater barriers to accessing primary care regularly \citep{caraballo2022trends}. Racial minorities also have lower health insurance rates \citep{healthcoverage2025}, an additional barrier to scheduling routine, preventative care visits. To evaluate the role of intake visits, we test whether removing the first timestamp associated with a patient’s DOB would eliminate reporting delays. If intake visits explained reporting delays, we would expect race information to be recorded by the second timestamp. While we observe a significant decrease in the percentage of patients with delays (\new{around} 20 percentage points), it does not appear that intake visits explain all of the reporting delays in our data.

\textit{Electronic health record (EHR) system-dependent lags:} Patients whose race is collected from practices with specific EHR systems might experience greater lags (\textit{e.g.}, due to specific data entry workflows). Prior research has shown tradeoffs between verbally collecting information from patients compared to paper forms or tablets \citep{american2013reducing}. Table~\ref{table:app-balance-table} provides some evidence of EHR-system differences.

Several features of our study may also limit the generalizability of some findings. We focus only on a primary healthcare setting, and do not include data from other external databases. Conclusions from research on hospitals and emergency departments may be less applicable. As mentioned earlier, the dataset and cohort used in our analyses were retrospective. In particular, the dataset was not specifically constructed to be a representative sample of the U.S. population, despite the AFC dataset having relatively representative coverage. Lastly, our study is limited to U.S. health data. Findings may differ in other geographic settings, or with other notions of social identity such as caste \cite{sambasivan2021re}.

\textbf{Imputation alone is not enough; practitioners should invest in improving data collection efforts on the ground.} We find that a widely used imputation method may perform satisfactorily on individual-level accuracy, but does not fundamentally improve group-level disparity assessments affected by reporting delays. 
In situations where timely feedback is necessary, our findings suggest practitioners should advocate for the suite of policy and programmatic changes that may reduce delays in the first place. It is important to consider principles such as data minimization~\cite{king2023privacy,biega2020operationalizing}, and ensure robust privacy protections~\cite{naiac2024} when designing strategies to incentivize timely data reporting.
Augmenting existing data collection with more reliable demographic data sources may be another promising direction --- \textit{e.g.}, integrating EHR and insurance data, which may have higher reporting standards.\looseness=-1

\textbf{Future work should study the impact of delays on other types of disparity metrics and ML-based metrics.} Importantly, outcomes of interest may not be binary (\textit{e.g.}, number of days to readmission after discharge from a hospital). 
Similarly, the impact of delayed race reporting should also be assessed in the context of fairness metrics corresponding to ML-based predictive models (\textit{e.g.}, models predicting clinical interventions such as vassopressor administration~\cite{ghassemi2015multivariate}). In this vein, our work connects to the literature on algorithmic audits in ML \citep{raji2020closing, lam2024framework} --- in particular, how missing and unreliable demographic data may impede auditing efforts \citep{ashurst2023fairness}. Our work suggests one reason why static audits -- completed once in time -- may fail to detect disparities. Furthermore, when conducting disparity assessments that involve aggregating data across sites or practices, it may be important to design data sampling and non-respondent follow-up strategies that account for delayed reporting.\looseness=-1

Lastly, it's worth noting that we exclude patients (approximately 11\%) for whom race information is \textit{never available}. Even though we are ultimately able to recover race data for all of the patients in our cohort --- those with reporting \textit{delays} --- traditional sources of missingness may also bias our disparity estimates.

\textbf{The existence of delays in race reporting underscores the importance of continuously and dynamically assessing fairness.}
Our results show that rates of missingness can be different between groups at different points of time. Hence, continuous assessment would be required to assess the robustness of conclusions made about disparities. It is important to consider sociotechnical systems as \emph{dynamic}, and how data missingness rates may change over time, driven by repeated user interactions with the same system (see Section~\ref{sec:continuous-monitoring} in the Appendix, where we analyze delayed reporting for consecutive patient cohorts and discuss implications for real-time monitoring scenarios). 
Prior work in algorithmic fairness has similarly raised the issue that fairness research should study the long-term impacts of deployed systems \citep{d2020fairness}. Our work aligns with such concerns, though we focus on underlying changes in the data. In particular, we urge fairness researchers to avoid treating data inputs as fixed, and to re-evaluate historical disparity assessments as more data that was initially missing becomes available over time. 
Our work also suggests another vulnerability to current static audit approaches: in the presence of reporting delays, providers might advertently or inadvertently leverage reporting delays to achieve more favorable audit outcomes.

\section{Conclusion}
In this work, we demonstrate the impact of delayed demographic information reporting when auditing the fairness of  decision-making systems. We focus on applications to healthcare where regular and timely monitoring of health disparities is critical. However, our work extends to any setting in which time-sensitive evaluations must be conducted prematurely. In a nationwide health dataset, we find that delayed reporting is a widespread problem, affecting nearly 3 out of every 4 patients. Furthermore, delays do not impact all patients evenly. Rates of delayed reporting vary by race and there are demographic, health, and practice-level differences between patients with and without delays. 
Furthermore, when we retrospectively estimate the impact of delays on disparity assessments, we find that delays can lead to inaccurate depictions of disparities. While these distortions are relatively small at the national level, there is greater heterogeneity for estimates at the state and practice levels --- an important consideration as recent health equity initiatives have occurred at the state level, and mitigation efforts necessarily start at the practice level. 

Broadly, our work highlights a crucial gap in the current auditing space: the need for frequent monitoring of systems reliant on seemingly ``static'' variables like race.
In fast-paced deployment environments, delays in specific data inputs may arise, leading to unexpected performance. 
Prior research has pointed out that access to individual-level demographic data is often unrealistic in real-world settings \citep{holstein2019improving, lee2021landscape, andrus2022demographic, andrus2021we, ashurst2023fairness}. Our work complicates this finding: demographic data can also be \emph{delayed}, thus highlighting an important direction for future work.

\section{Ethical Statement}

We now address ethical considerations that arose in the course of this work. 

First, our work deals with sensitive patient health data. All analyses were conducted on secure, remote servers approved for High Risk and Protected Health Information (PHI) data, and the research was approved by the Institutional Review Board, including Waiver of Informed Consent, Waiver of Assent, and Waiver of HIPAA Authorization. To protect confidentiality, we only produced aggregated results for cell sizes $> 10$. The American Family Cohort dataset is used solely for research purposes, allowing researchers to investigate core questions of health equity and to generate knowledge that may inform the improvement of healthcare services across a nationwide network of primary care practices.

Second, another ethical consideration lies in the measurement of race. Through all of our findings, our central focus is on the sobering reality that health disparities exist between different groups within communities across the U.S. The true nature of these disparities are, of course, always more complex and intersectional than the socially constructed racial categories we choose to use at any given moment (\textit{i.e.}, individuals' self-reported racial categories may not align with federal categories or may change over time), and the improper reification of racial categories (including through statistical imputation methods such as BIFSG) may itself run the risk of feeding back into the entrenchment or exacerbation of those disparities. At the same time, in the absence of any records of race identification, we may not be able to detect and act upon disparities at all. Central to algorithmic fairness has been the notion of ``fairness through awareness''~\citep{dwork2012fairness}. At core, our study identifies an underappreciated mechanism, delayed demographic reporting, by which that awareness can be obfuscated. Critically, the utilization of racial categories, obtained through self-reporting or through imputation methods, to \textit{assess disparities} does not in any way imply that they should also be used in individual-level medical decision making \cite{ho2020affirmative}, where the ethical considerations may be significantly more acute.

\begin{acks}
We thank Nathaniel Hendrix, Shiying Hao, Esther Velásquez, Ayin Vala, and Isabella Chu for their support with using American Family Cohort data, and Julian Morimoto, Claire Morton, and Jenn Wang for their helpful feedback. We are grateful to the Stanford Institute for Human-Centered Artificial Intelligence and the Robert Wood Johnson Foundation Evidence 4 Action Program for supporting this research.
\end{acks}

\bibliographystyle{ACM-Reference-Format}
\bibliography{references.bib}

\clearpage
\appendix

\section{Detecting when Race and Ethnicity is Unknown or Declined}
\label{sec:unknown_race_parsing}
The dataset contains multiple datapoints (at different timepoints) per patient / practice. Each datapoint is associated with a modification date of race. Further, each datapoint contains two free-text fields -- `patientracetext' and `patientethnicitytext'. Race codes are also available in fields of `patientracecode' and `patientraceethnicity'. We define patient race to be unknown or declined at a given timepoint if both the following conditions are met:
\begin{enumerate}
    \item Free text fields (after lower-casing) are in the following list:  `race not reported - don't know', `nh', `unspecified', `do not use', `not hispanic/latino ethnicity', `other/declined', `non-hispanic', `not of hispanic, latino/a or spanish origin', `refuse', `not hispanic', `declined', `unknown', `patient declined information', `unreported/refused to report', `patient declined', `not set', `refuse to report/ unreported', `declined to answer', `not reported', `non hispanic', `withheld', `unknown/unwilling', `u', `<none>', `NA', `unknown/unreported', `decline to answer', `refused to report', `unknown / not reported', `not hispanic or latino', `non hispanic-non latino', `non - hispanic/latino', `prefers not to answer', `*unspecified', `refused', `refused to report/unreported', `unknown/not reported', `not hispanic / latino', `dec', `refused, unknown', `undefined', `chose not to disclose', `unk', `race not reported - refusal', `non hispanic or latino', `n', `nsp', `x', `unk', `declines to state', `unavailable / unknown', `refus', `dec', `not hispanic, latino/a, or spanish origin', `non-hispanic / non latino', `state prohibited', `decline', `declined to specify', `not provided', `patient refused', `un', `unreported / refused to report', `race not reported - not ascertained', `unknown to patient', `declines to specify', `decli', `dc', `ds', `ua', `uo', `n', `d', `u', `r', `unkno', `nr', `unreported / unknown (uds)', `unreported / unknown', `unavailable', `2186-5', `9'. 
    \item Categorical codes are either invalid codes or strings indicating no information. Specifically, they are among the following list: `UNK', `NA', `UN', `U', `2186-5', nan, `UNK', `UN', `2186 - 5', `N', `312507', `NH', `NR', `ASKU'. 
\end{enumerate}
 Note that if some information is provided in either field -- race or ethnicity -- we do not consider race to be \new{unreported}. The only exception is when patients only report that they are non-Hispanic, with no other race information provided. In such cases, race is considered missing or delayed.

\section{Extraction of Health Outcomes}
\label{sec:extract_health_outcomes}
We extract six clinical outcomes curated based a literature review: \textit{i.e.}, there is evidence of disparities between
racial group for each of these outcomes. 
For example, prior work has documented higher depression screening rates among Black and Asian patients compared to White patients~\cite{hahm2015intersection}.  Health outcomes in each case are extracted by relying primarily on Systematized Nomenclature of Medicine (SNOMED) codes, which are used for clinical documentation and billing purposes. Codes in each case are retrieved using a database search tool. We extract health outcomes based on the presence of specific diagnosis, documentation, and billing codes. We rely primarily on SNOMED-CT codes because they are also used for clinical documentation, apart from just billing. To identify relevant codes per outcome, we use the Athena search tool.\footnote{\url{https://athena.ohdsi.org/search-terms/start}} Using a relevant search term per outcome, we retrieve a list of codes that are returned as matches per search. Then, we filter these codes manually by reading the text description corresponding to the code. We assume that a code, if entered in the system, is accurate. An overview of the codes for each outcome is provided in Table~\ref{tab:snomed_codes}. To fully account for outcomes, we also include codes where the presence of an outcome is indirectly indicated. For example, the SNOMED-CT code corresponding to the condition of ``Senile dementia with depression'' is also included in the list of codes for identifying the outcome of depression diagnosis.

\begin{table}[H]\centering
\scriptsize
\begin{tabular}{lrrr}\toprule
\textbf{Clinical Outcome} &\textbf{Vocabulary} &\textbf{Number of codes (example)} \\\midrule
Depression diagnosis &SNOMED, OMOP Extension &104 (e.g., 35489007) \\
Diabetes diagnosis &SNOMED, OMOP Extension &92 (e.g., 771000119108) \\
Hypertension diagnosis &SNOMED, OMOP Extension &10 (e.g., 78975002) \\
Diabetes HBA1c measurement &SNOMED, OMOP Extension &1 (43396009) \\
Electrocardiogram procedure &SNOMED, CPT4, HCPCS &38 (e.g., 93005) \\
Depression screening procedure &SNOMED, CPT4, HCPCS &4 (e.g., 96127) \\
\bottomrule
\end{tabular}
\caption{Overview of codes for each outcome variable. }\label{tab:snomed_codes}
\end{table}

\section{Cohort Size versus Delay}
\begin{figure}[htb!]
\centering
\includegraphics[width=\linewidth]{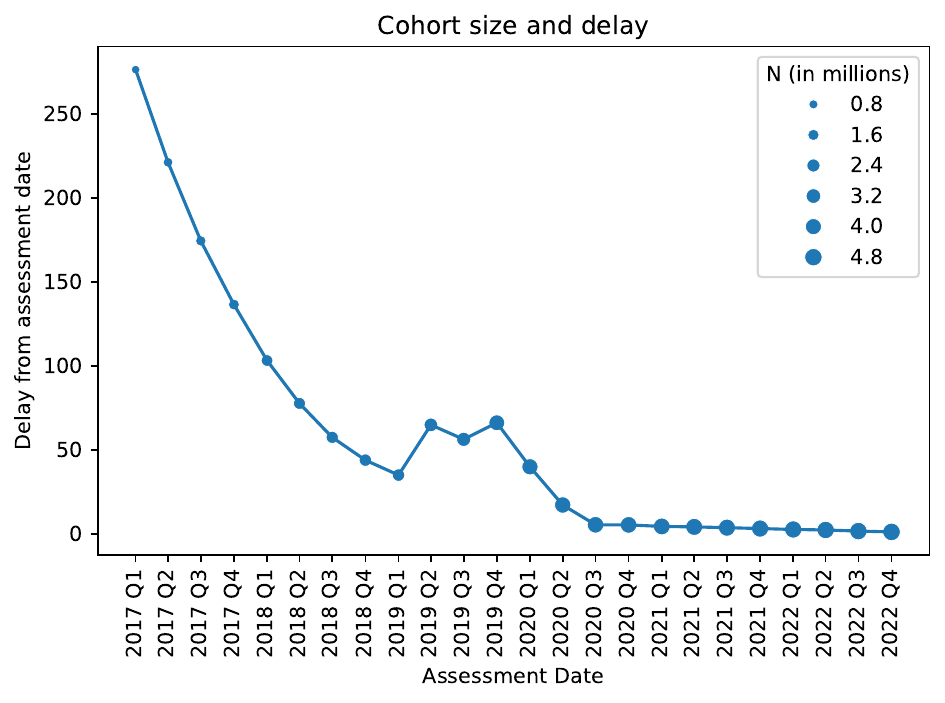}
\caption{Cohort size vs average delay of race reporting for different quarters as the focus of disparity assessment.
\label{fig:appendix_cohort_size}}
\Description{Cohort sizes increase while average delay decreases over time. On the x-axis, the cohort date is shown. Average delay in days is shown on the y-axis. The total cohort size in millions is represented by the size of the marker.}
\end{figure}

In Figure~\ref{fig:appendix_cohort_size}, we visualize cohort size versus average delay (in days), as they vary across different quarters for which to conduct the disparity assessment. We choose the 2018 Q1 cohort because it has a reasonable sample size, as well as a high average delay.

\section{Comparing Patients with and without Delays}
Table~\ref{table:app-balance-table} shows the average difference in demographic characteristics and all other variables considered in the main text.
\begin{table*}
\centering
\begin{tabular}{p{0.65in}p{1.5in}S|SSS}
& &  \textbf{Overall Average} &  \textbf{No Delay} &  \textbf{Delay} &  \textbf{Difference} \\
\toprule
 & N & \text{5,310,700} & \text{1,399,487} & \text{3,911,213} & \\
 \hline
\multicolumn{6}{c}{Patient-level Characteristics} \\
\hline
\multirow{12}{*}{Demographics}
& Age (years) & 58.02 & 55.97 & 58.76 & 2.79*** \\
& Female (\%) & 56.26 & 55.34 & 56.59 & 1.25*** \\
& Male (\%) & 43.69 & 44.58 & 43.37 & -1.21*** \\
& Other (\%) & 4.69 & 7.61 & 3.65 & -3.96*** \\
& AIAN (\%) & 0.73 & 0.73 & 0.74 & 0.01 \\
& Asian (\%) & 2.69 & 2.78 & 2.65 & -0.12*** \\
& Black (\%) & 8.45 & 8.78 & 8.33 & -0.45*** \\
& NHPI (\%) & 0.52 & 0.66 & 0.47 & -0.18*** \\
& White (\%) & 80.25 & 77.49 & 81.24 & 3.75*** \\
& Hisp (\%) & 10.91 & 13.52 & 9.98 & -3.54*** \\
& Other (\%) & 4.12 & 4.78 & 3.89 & -0.89*** \\
& Multi (\%) & 0.89 & 0.93 & 0.87 & -0.07*** \\
\hline
\multirow{4}{*}{Marital Status} 
& Single (\%) & 28.28 & 30.26 & 27.64 & -2.62*** \\
& Married (\%) & 58.72 & 57.26 & 59.19 & 1.93*** \\
& Divorced (\%) & 7.42 & 7.33 & 7.45 & 0.12*** \\
& Widowed (\%) & 5.51 & 5.02 & 5.67 & 0.65*** \\
& Partner (\%) & 0.07 & 0.12 & 0.05 & -0.08*** \\
& Other (\%) & 4.69 & 7.61 & 3.65 & -3.96*** \\
\hline
\multirow{2}{*}{Visits} 
& Avg. yearly visits (2007-2023) & 0.82 & 0.50 & 0.93 & 0.43*** \\
& Earliest year & 2016.23 & 2017.08 & 2015.92 & -1.15*** \\
\hline
\multirow{6}{*}{Health} 
& Diabetes (\%) & 10.84 & 8.54 & 11.66 & 3.12*** \\
& Depression (\%) & 12.33 & 9.90 & 13.20 & 3.30*** \\
& Hypertension (\%) & 25.17 & 20.24 & 26.93 & 6.70*** \\
& Depression screen (\%) & 14.44 & 10.26 & 15.94 & 5.67*** \\
& Electrocardiogram (\%) & 26.84 & 18.63 & 29.78 & 11.15*** \\
& Hba1c (\%) & 25.46 & 17.74 & 28.22 & 10.49*** \\
\hline
\multicolumn{6}{c}{Practice-level Characteristics (Mapped to Patients)}\\
\hline
\multirow{2}{*}{Practice Info.} 
& Available (\%) & 85.93 & 86.25 & 85.82 & -0.43*** \\
& Unavailable (\%) & 14.07 & 13.75 & 14.18 & 0.43*** \\
\hline
\multirow{2}{*}{Data Source} 
& EHR \& PM (\%) & 6.38 & 9.86 & 5.13 & -4.73*** \\
& EHR only (\%) & 93.02 & 89.59 & 94.25 & 4.66*** \\
\hline
\multirow{3}{*}{Data Update}
& Push, pull, and other (\%) & 17.43 & 21.87 & 15.84 & -6.03*** \\
& Pull only (\%) & 81.07 & 76.39 & 82.76 & 6.37*** \\
& Other (\%) & 1.45 & 1.70 & 1.36 & -0.34*** \\
\hline
\multirow{11}{*}{EHR System}
& Multiple (\%) & 17.71 & 22.16 & 16.11 & -6.05*** \\
& eMDs - Solution Series (\%) & 51.96 & 45.19 & 54.40 & 9.21*** \\
& Amazing Charts (\%) & 8.60 & 9.22 & 8.38 & -0.84*** \\
& eMDs - Practice Partner (\%) & 3.18 & 4.43 & 2.73 & -1.70*** \\
& Veradigm EHR (\%) & 3.00 & 3.67 & 2.76 & -0.91*** \\
& eClinicalWorks (\%) & 1.82 & 1.12 & 2.07 & 0.95*** \\
& GE Centricity (\%) & 1.75 & 1.20 & 1.95 & 0.75*** \\
& Aprima (\%) & 1.63 & 1.55 & 1.66 & 0.11*** \\
& Athenahealth (\%) & 1.33 & 2.06 & 1.07 & -0.99*** \\
& eMDs - Lytec MD (\%) & 1.21 & 2.02 & 0.91 & -1.11*** \\
& Other (\%) & 7.22 & 7.00 & 7.30 & 0.30*** \\
\hline
\end{tabular}
\caption{Differences in average demographic characteristics, visits, health outcomes, and practice-level characteristics between patients who experience delays compared to patients with no delays. Most of the differences are statistically significant with $***=p<0.01$ even with Benjamini-Hochberg adjustment for multiple comparisons. On average, patients who experience delays are older, more likely to be White, and have more visits. They also have higher prevalence rates of adverse health conditions and procedures. AIAN = American Indian or Alaska Native; NHPI = Native Hawaiian or Pacific Islander; EHR = electronic health record; PM = practice management system.}
\label{table:app-balance-table}
\end{table*}

\section{Retrospective Analysis for Additional Health Outcomes}
\label{app:additional-outcomes}

\begin{figure*}
    \centering
    \includegraphics[width=0.8\linewidth]{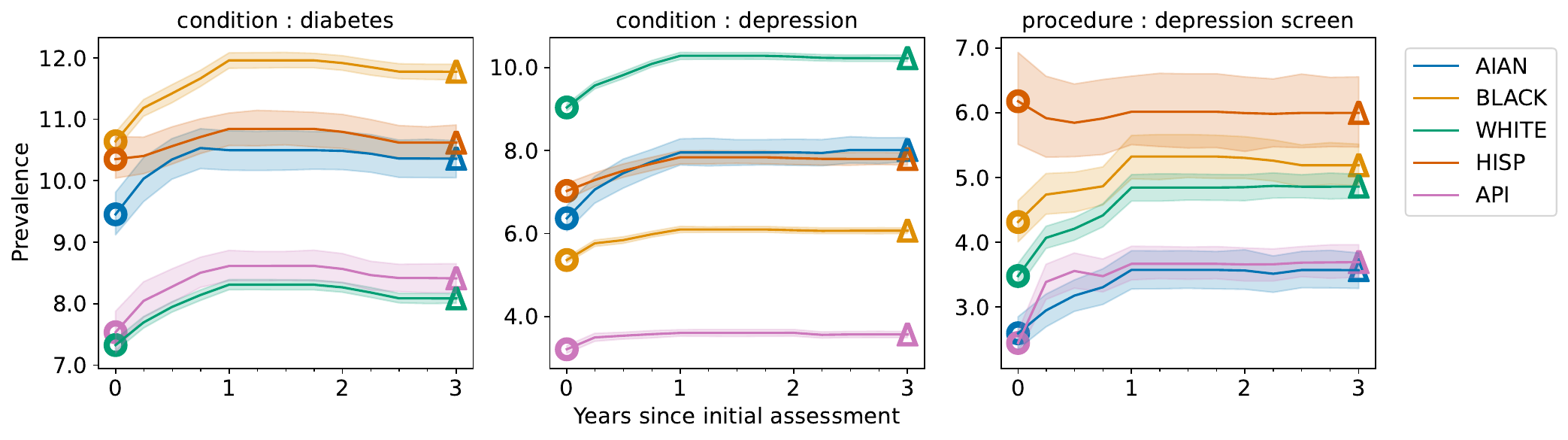}
    \caption{Simulations at the national level for two conditions (diabetes and depression) and one procedure (depression screens). All disparity estimates are conducted on a fixed cohort of patients from 2018 Q1.}
    \label{fig:simulation-appendix}
    \Description{Prevalence estimates at the national level change over time due to delays in race reporting. On the x-axis, the number of years from the initial assessment date (2018 Q1) is shown. Prevalence estimates for each health outcome disaggregated by race  appear on the y-axis. 95\% confidence intervals represent uncertainty. In general, prevalence estimates for most race groups increase over time as more race information is recovered.}
\end{figure*}

\begin{figure*}
    \centering
    \includegraphics[width=0.8\linewidth]{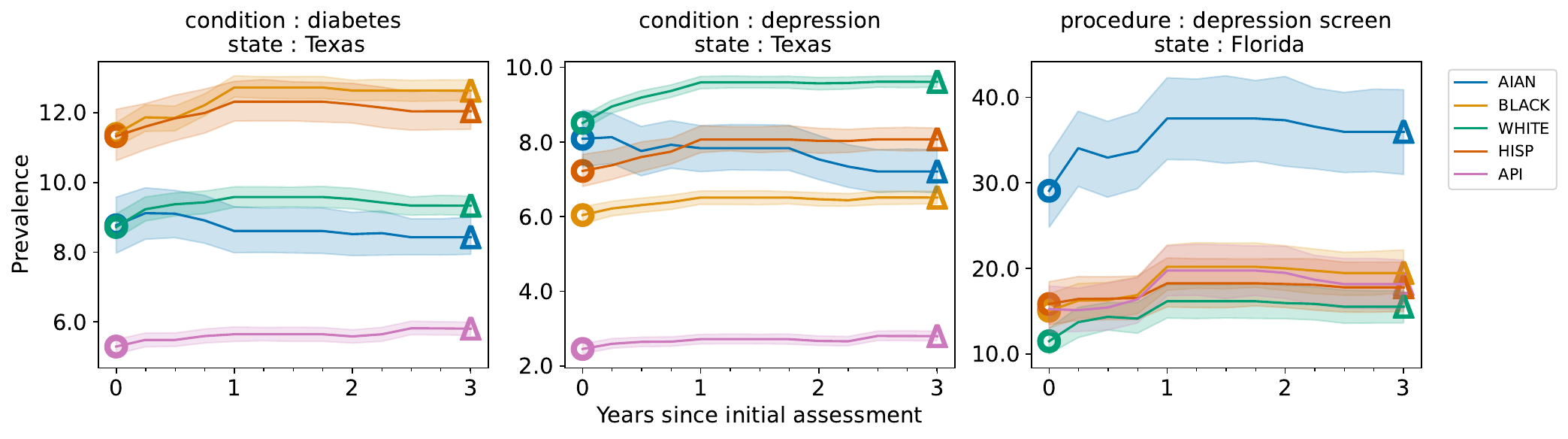}
    \caption{Simulations at the state level for two conditions (diabetes and depression) and one procedure (depression screens). The states shown are among the top three states with the highest average number of patients with each health outcome. All disparity estimates are conducted on a fixed cohort of patients from 2018 Q1.}
    \label{fig:simulation-state-appendix}
    \Description{Prevalence estimates at the state level (for Texas and Florida) change over time due to delays in race reporting. On the x-axis, the number of years from the initial assessment date (2018 Q1) is shown. Prevalence estimates for each health outcome disaggregated by race appear on the y-axis. 95\% confidence intervals represent uncertainty. In general, prevalence estimates for most race groups increase over time as more race information is recovered.}
\end{figure*}

Figure~\ref{fig:simulation-appendix} and Figure~\ref{fig:simulation-state-appendix} show the results from the retrospective analysis at both the national and state level for three additional health outcomes: diabetes, depression, and depression screens. 

\section{Delayed Reporting for 2017 Q1 and 2018 Q1 Cohorts}
\label{sec:other-cohorts}

\begin{figure*}
\centering
\begin{subfigure}[t]{\textwidth}
\centering
    \includegraphics[width=0.8\linewidth]{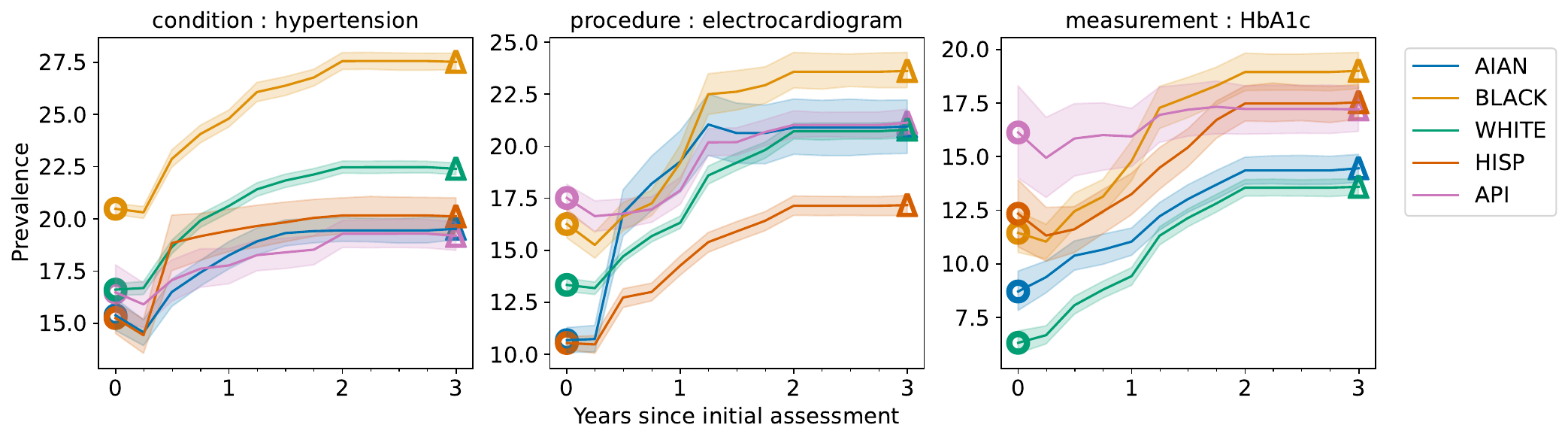}
\end{subfigure}
\begin{subfigure}[t]{\textwidth}
\centering
    \includegraphics[width=0.8\linewidth]{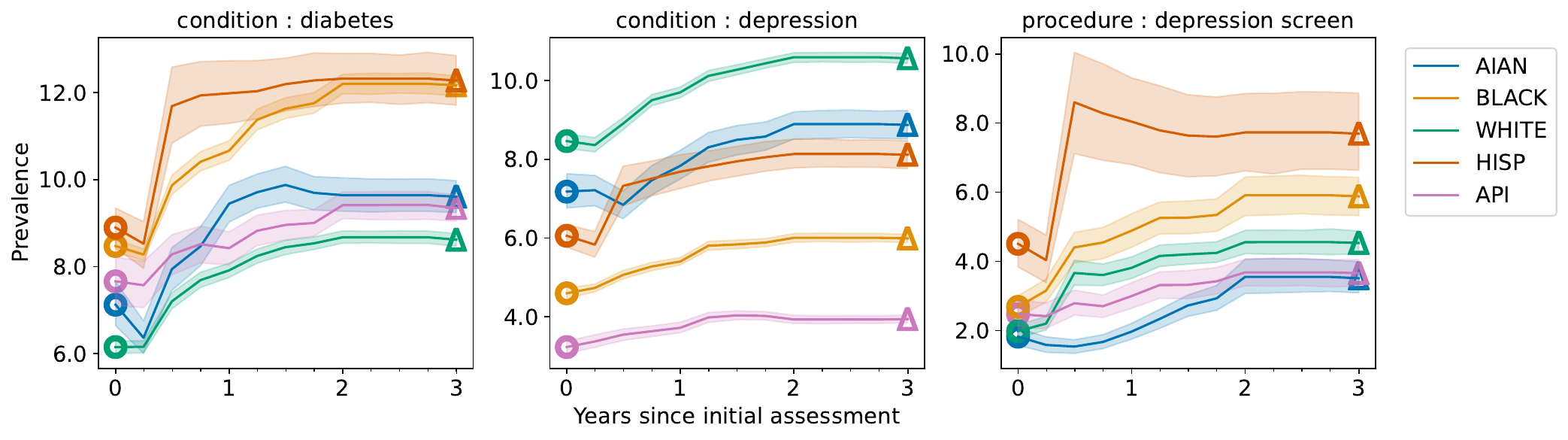}
\end{subfigure}
\caption{\new{Simulations at the national level for a fixed cohort of patients from 2017 Q1.}}
\label{fig:appendix-2017-cohort}
\Description{Prevalence estimates at the national level for a cohort of patients from 2017 Q1. On the x-axis, the number of years from the initial assessment date (2017 Q1) is shown. Prevalence estimates for each health outcome disaggregated by race  appear on the y-axis. 95\% confidence intervals represent uncertainty. In general, prevalence estimates for most race groups increase over time as more race information is recovered. Increases over time are more striking compared to the cohort from 2018 Q1.}
\end{figure*}

\begin{figure*}[htb!]
\centering
\begin{subfigure}[t]{\textwidth}
\centering
    \includegraphics[width=0.8\linewidth]{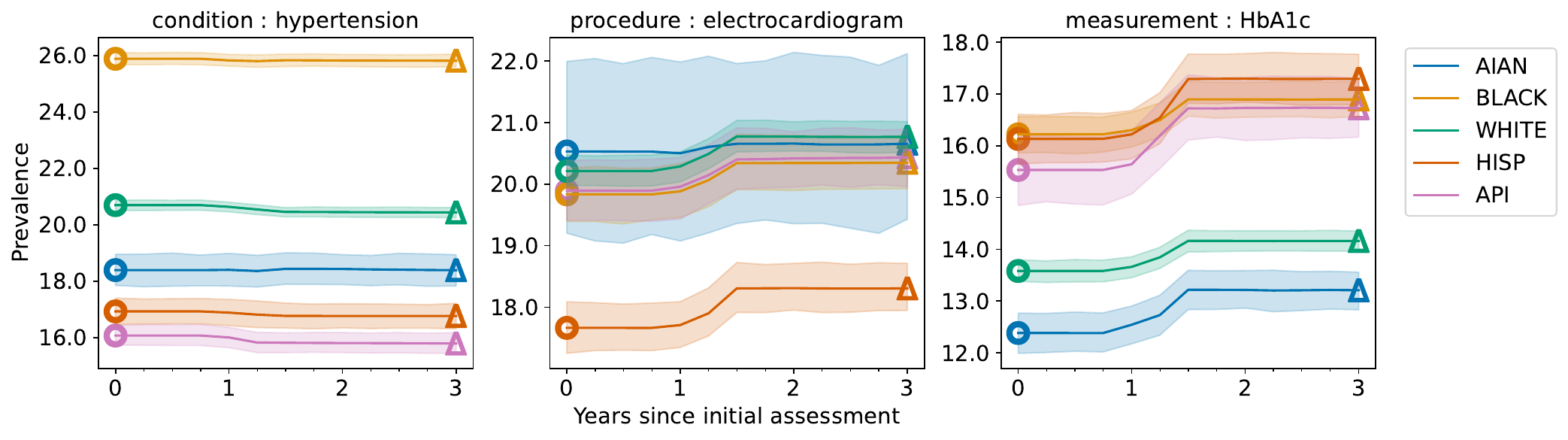}
\end{subfigure}
\begin{subfigure}[t]{\textwidth}
\centering
    \includegraphics[width=0.8\linewidth]{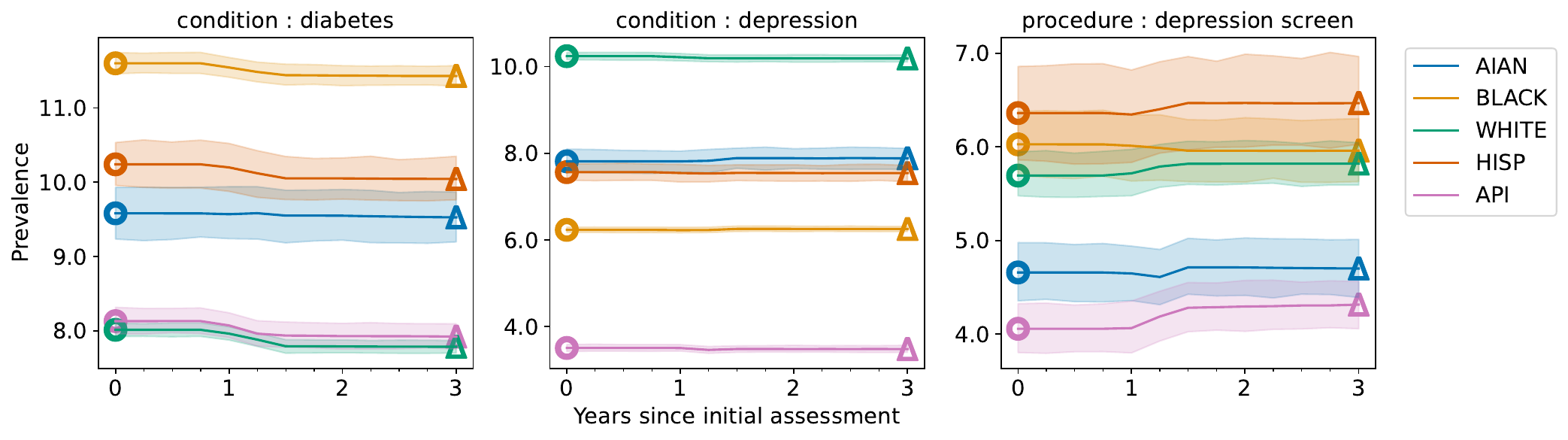}
\end{subfigure}
\caption{\new{Simulations at the national level for a fixed cohort of patients from 2019 Q1.}}
\label{fig:appendix-2019-cohort}
\Description{Prevalence estimates at the national level for a cohort of patients from 2019 Q1. On the x-axis, the number of years from the initial assessment date (2019 Q1) is shown. Prevalence estimates for each health outcome disaggregated by race  appear on the y-axis. 95\% confidence intervals represent uncertainty. In general, prevalence estimates for most race groups remain consistent over time even though more race information is recovered. Changes over time are smaller compared to the cohort from 2018 Q1.}
\end{figure*}

Figures~\ref{fig:appendix-2017-cohort} and \ref{fig:appendix-2019-cohort} implement national-level disparity assessments for different cohorts---2017 Q1 and 2019 Q1. They show temporal differences in the effect of delays. In 2019 Q1, delays play a minor role---affecting prevalence and disparity estimates slightly (\textit{e.g.}, the Black-API disparity for HbA1c measurements decreases in $t_{\text{3Yrs}}$) compared to $t_{\text{initial}}$). Delays have a much larger effect on disparity estimates for the 2017 Q1 cohort in line with the findings from Figure~\ref{fig:appendix_cohort_size} that show greater average delay in 2017 compared to 2019 and later.

\section{Relative Absolute Prevalence Error}
\label{sec:relative-absolute-error}

We show relative absolute prevalence error by race in Table~\ref{tab:relative-absolute-error}. Relative absolute error accounts for class imbalance, and shows the magnitude of the prevalence error relative to the true overall prevalence. For example, prevalence errors for hypertension and depression are broadly similar among AIAN, Hispanic, and White subgroups in Table~\ref{tab:prevalence-rates}. But relative absolute errors (shown here) are much larger for depression compared to hypertension since the true overall prevalence of depression overall is smaller.

\begin{table}
\centering
\begin{tabular}{lSSSSS}
   {\textbf{Health outcome}} & \multicolumn{5}{c}{\textbf{Relative absolute prevalence error}}  \\
   \cline{2-6}
 & \text{AIAN} & \text{API} & \text{Black} & \text{Hispanic} & \text{White} \\
  \hline

Diabetes & 8.79 & 10.79 & 9.54 & 3.92 & 9.34 \\
Hypertension & 8.62 & 7.72 & 7.87 & 3.81 & 8.87 \\
Depression & 20.60 & 10.00 & 11.63 & 9.97 & 11.57 \\
Depression screen & 27.47 & 33.28 & 17.70 & 10.45 & 28.67 \\
Electrocardiogram & 8.16 & 23.53 & 21.92 & 18.74 & 23.11 \\
HbA1c & 33.39 & 22.26 & 31.11 & 33.02 & 38.24 \\

\hline
\end{tabular}
\caption{\new{Relative absolute errors in prevalence between $t_\text{initial}$ and $t_\text{final}$ at the national level for each racial group and health outcome, focused on health outcomes for a cohort in 2018 Q1. Values provided are percentage points.}}
\label{tab:relative-absolute-error}
\vspace{-1em}
\end{table}

\section{Distribution of Errors in Prevalence}
\begin{figure}[htbp!]
        \centering
        \includegraphics[width=\linewidth]{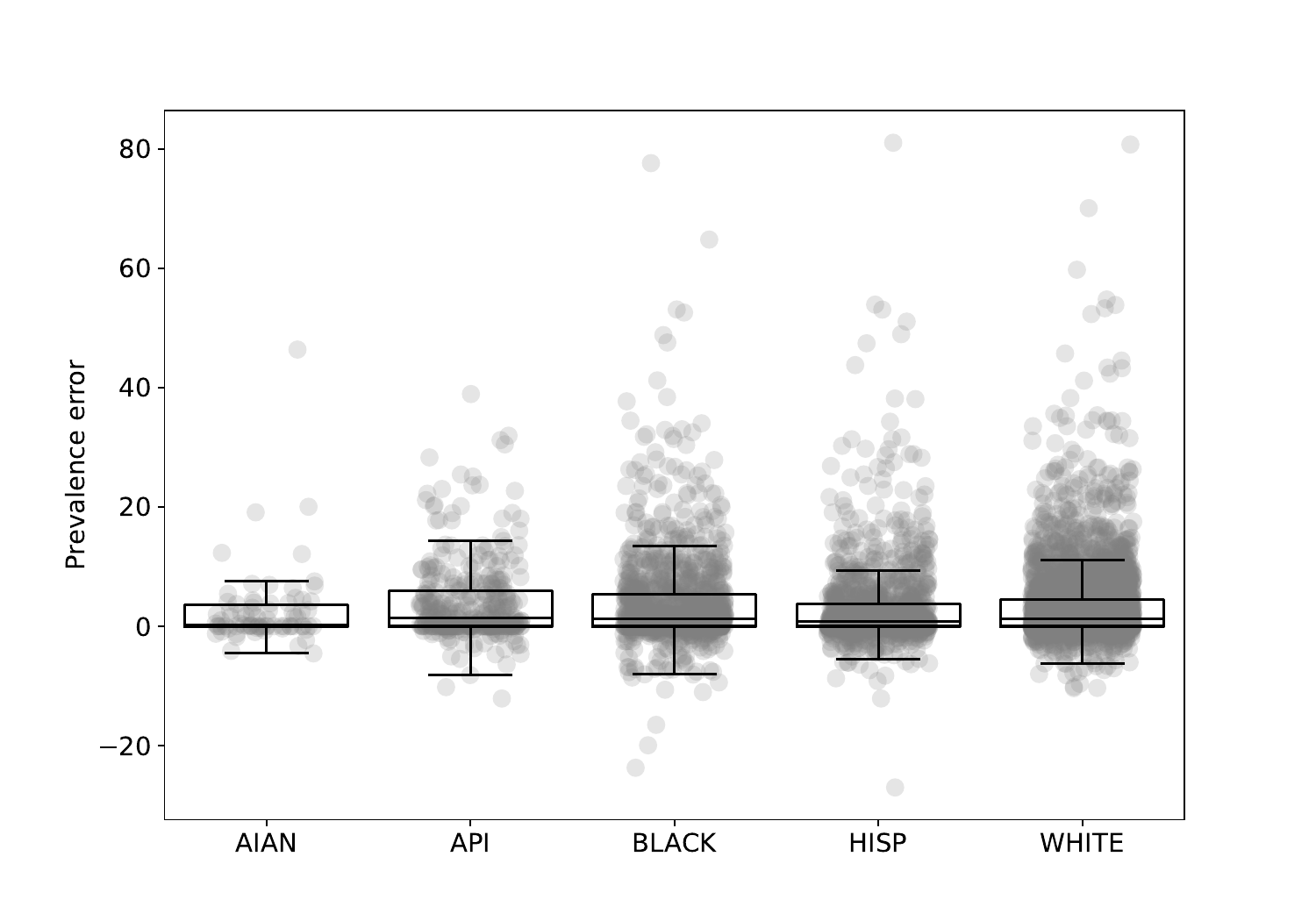} 
    \caption{Distribution of prevalence errors for all conditions averaged across all bootstrapped samples at the practice level. We remove all practice-level prevalence estimates that involve fewer than $10$ patients on average to preserve patient privacy.}
    \label{fig:practice-dist-differences}
    \Description{A boxplot and scatter plot for each race group shows the distribution of prevalence error at the practice level. Race groups are shown on the x-axis and prevalence error on the y-axis.}
\end{figure}

Figure~\ref{fig:practice-dist-differences} shows the distribution of prevalence errors at the practice level across different racial groups and for all health outcomes. We present prevalence errors averaged across $50$ bootstrapped samples. Most of the errors are centered around 0, though there are numerous outliers. Overall, prevalence errors skew positive, reflecting a trend similar to the national level (Table~\ref{tab:prevalence-rates}).

\section{Exacerbation, Minimization, and Sign Switches in Disparity Assessment Error}
\label{app:ex_min_sign}
We also consider whether delays consistently lead to over- or under-estimates of the true disparity in absolute terms or whether the direction of the disparity changes entirely. We call the case of an over-estimate an \emph{exacerbation} and the case of an under-estimate a \emph{minimization}. When we observe exacerbation, the disparity at $t_\text{initial}$ appears higher than it actually is at $t_{\text{final}}$ (and in minimization, lower than it actually is). Exacerbations may be a concern when there are limited resources, but the consequences for not intervening are small. Minimizations may be a concern when there are serious health concerns that would require immediate intervention. \emph{Sign changes} (\textit{i.e.}, the direction of the disparity changes, which can be either exacerbations or minimizations) are a problem in either setting as they would distort any meaningful takeaways related to the disparity. 

Table~\ref{tab:simulation-summary} summarizes these trends by comparing \textit{disparity estimates} (\textit{i.e.}, pairwise differences in prevalence across groups, for each outcome) between $t_\text{initial}$ and $t_{\text{final}}$, at the national level. %
Delays are more likely to minimize the true disparity for three outcomes; \textit{i.e.}, premature evaluations often underestimate disparities. But delays are more likely to exacerbate the true disparity for the other three outcomes; \textit{i.e.}, premature evaluations also often overestimate disparities. Perhaps most concerning is the high rate of sign switching for an outcome like electrocardiogram procedures; \textit{i.e.}, premature evaluations can be wrong about the direction of disparity on average $14\%$ of the time. %

\begin{table}[!ht]
\centering
\resizebox{\linewidth}{!}{%

\begin{tabular}{lccc}
\toprule
  & \textbf{\%Exacerbation} & \textbf{\%Minimization} & \textbf{\%Sign Switch}
\\\midrule
Diabetes & 51.6 & 41.2 & 7.2 \\
Depression & 21.2 & 73.8 & 5.0 \\
Hypertension & 29.4 & 64.6 & 6.0 \\
Depression screen & 56.4 & 32.8 & 10.8 \\
Electrocardiogram & 40.0 & 46.0 & 14.0 \\
HbA1c & 43.4 & 42.8 & 13.8 \\

\hline
\end{tabular}
}
\caption{Comparison of errors in disparity estimation. We classify sign switches (\textit{i.e.}, rank order changes for the pairwise comparisons, which can be either exacerbations or minimizations as well) first, and then classify the remaining errors as exacerbations or minimizations.}
\label{tab:simulation-summary}
\end{table}

\section{BIFSG: Impact on Error Metrics}

\begin{figure}[htb!]
\centering
\includegraphics[width=0.8\linewidth]{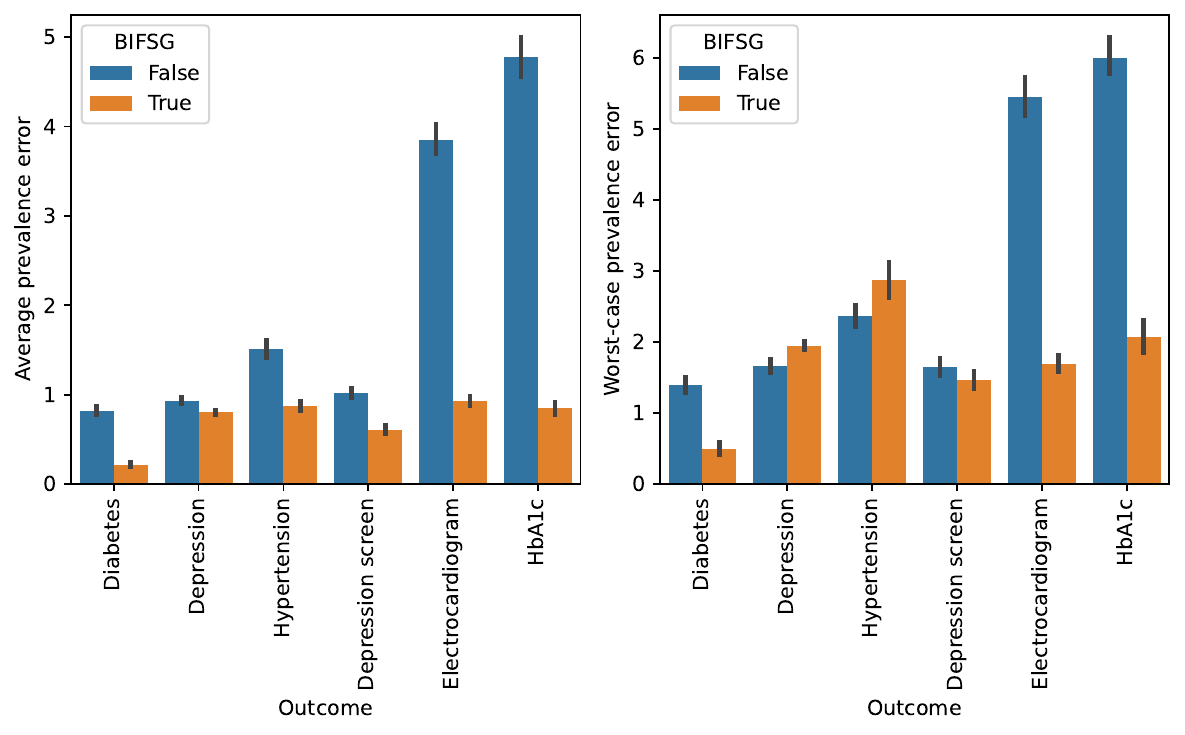}
\caption{BIFSG reduces the average prevalence error for all outcomes and the worst-case prevalence error for most outcomes.
\label{fig:appendix_bifsg_impute}}
\Description{There are two plots shown in this figure. On the left, we show the average prevalence error for estimates with and without BIFSG. On the right, we show the worst-case prevalence error for estimates with and without BIFSG. The six health outcomes are shown on the x-axis (Diabetes, Depression, Hypertension, Depression screening, ECG, and HbA1c). Blue bars represent errors for estimates produced without BIFSG, and orange bars represent errors for estimates produced with BIFSG. For most outcomes, estimates produced with BIFSG have lower average and worst-case prevalence error.}
\end{figure}

\begin{figure}[htb!]
\centering
\includegraphics[width=0.5\linewidth]{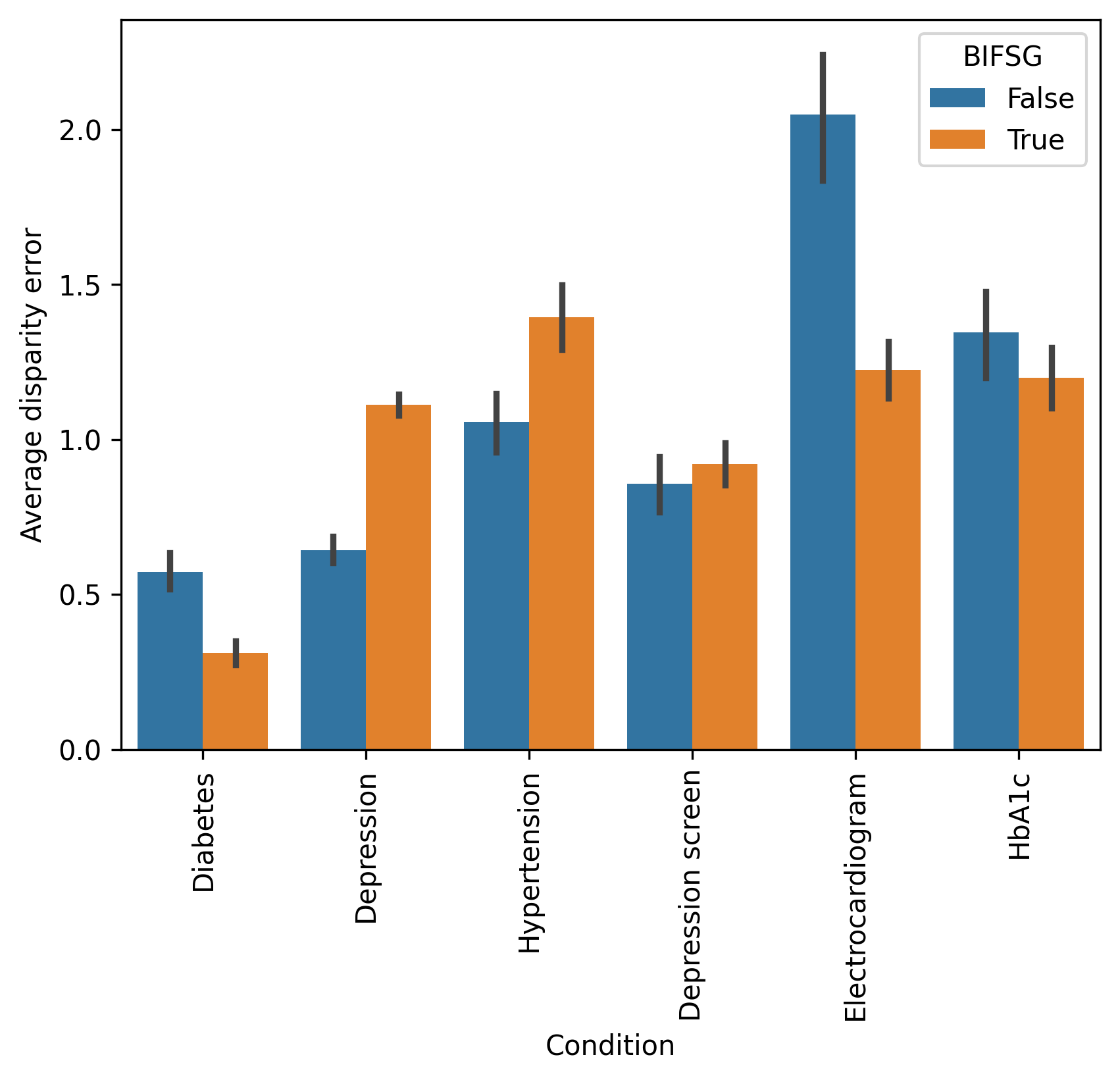}
\caption{BIFSG mitigates error in disparity assessment (disparity error) in three out of six outcomes, but not the others.
\label{fig:appendix_bifsg_impute2}}
\Description{On the x-axis, the six health outcomes are shown (Diabetes, Depression, Hypertension, Depression screening, ECG, and HbA1c). Average disparity error is shown on the y-axis. Blue bars represent average disparity errors for estimates produced without BIFSG, and orange bars represent average disparity errors for estimates produced with BIFSG. The average disparity errors are higher for estimates produced with BIFSG (orange bars) in three out of the six outcomes.}
\end{figure}

\begin{figure*}
\centering
\includegraphics[width=0.95\linewidth]{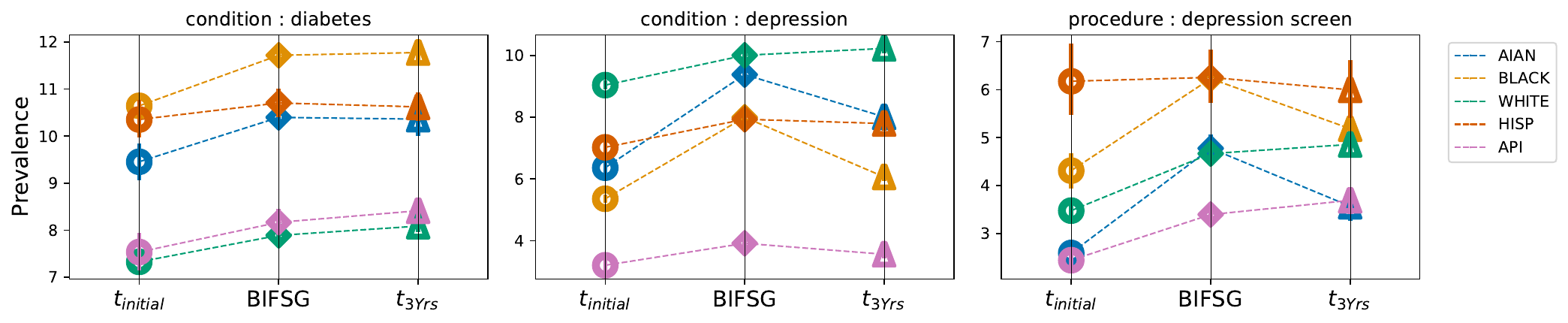}
\caption{BIFSG over-estimates prevalence for several minority race groups (\textit{e.g.}, Black and AIAN patients) across several outcomes, though average prevalence error is improved. In each subplot, the y-axis denotes the estimated prevalence. Values for $t_{initial}$ and $t_{3Yrs}$ match the same national values as shown in Figure~\ref{fig:simulation-appendix}.}
\label{fig:bifsg_impute_all}
\Description{BIFSG estimates in comparison to estimates at $t_{\text{initial}}$ and $t_{\text{3Yrs}}$ for three additional health outcomes (diabetes, depression, and depression screenings). We show all three types of estimates on the x-axis in order ($t_{\text{initial}}$, BIFSG, and $t_{\text{3Yrs}})$. Prevalence is shown on the y-axis.}
\end{figure*}

BIFSG improves average prevalence error (Figure~\ref{fig:appendix_bifsg_impute}), but the direction of error changes in some cases (Figures~\ref{fig:bifsg_impute} and~\ref{fig:bifsg_impute_all}). We also visualize worst-case prevalence error, or the highest absolute gap in prevalence error across groups. Furthermore, BIFSG mitigates disparity error in three out of six outcomes (see Figure~\ref{fig:appendix_bifsg_impute2}).

\section{Delayed Reporting across Consecutive Disparity Assessments}
\label{sec:continuous-monitoring}

\begin{figure*}
\centering
\begin{subfigure}[t]{0.3\textwidth}
\includegraphics[width=\linewidth]{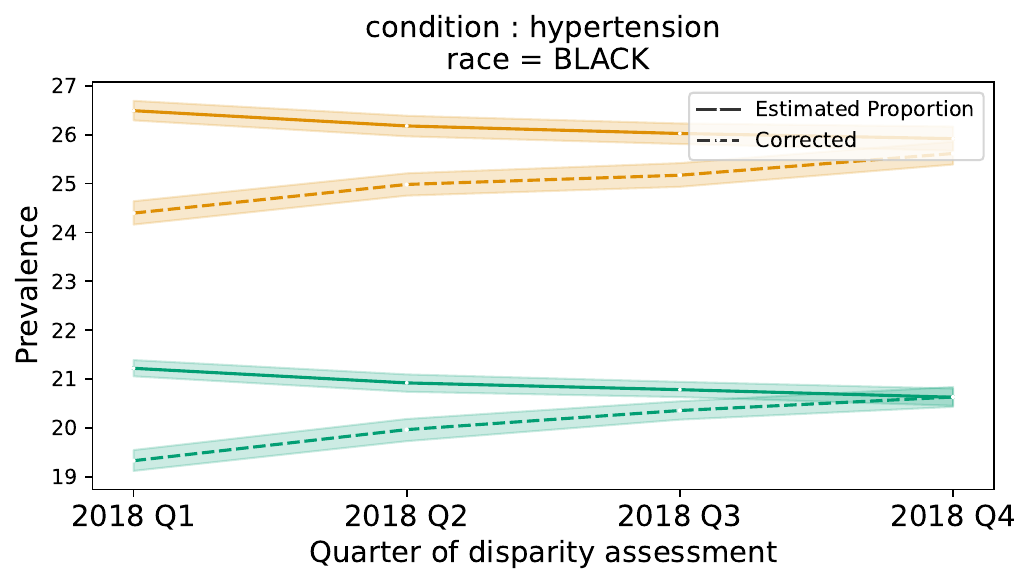}
\end{subfigure}
\begin{subfigure}[t]{0.3\textwidth}
\includegraphics[width=\linewidth]{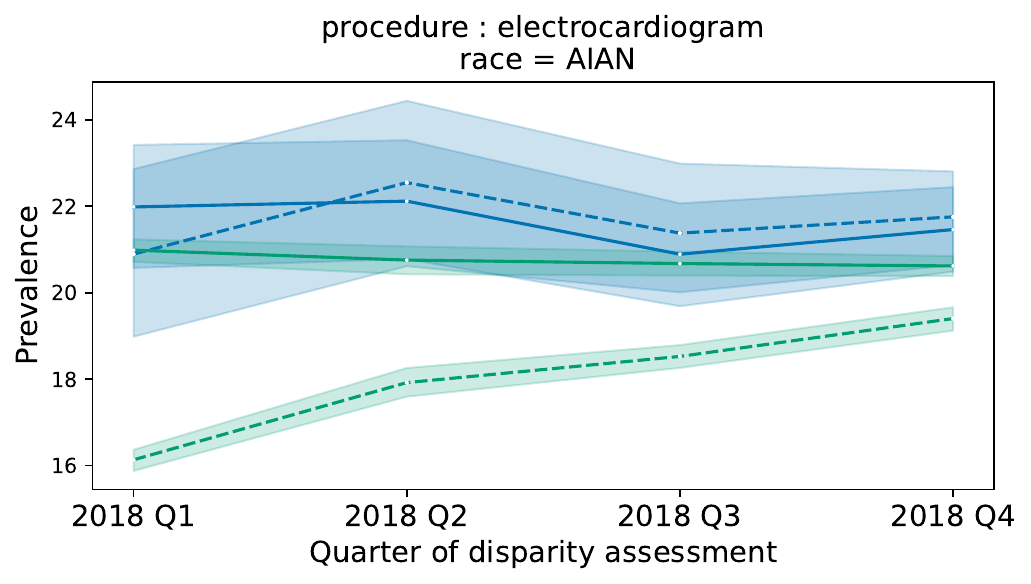}
\end{subfigure}
\begin{subfigure}[t]{0.3\textwidth}
\includegraphics[width=\linewidth]{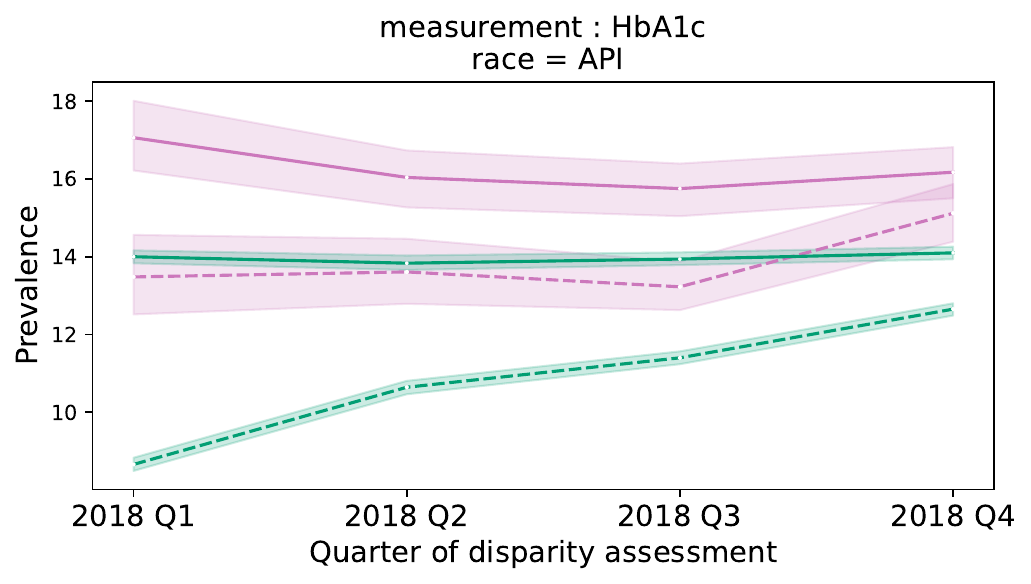}
\end{subfigure}
\begin{subfigure}[t]{0.3\textwidth}
\includegraphics[width=\linewidth]{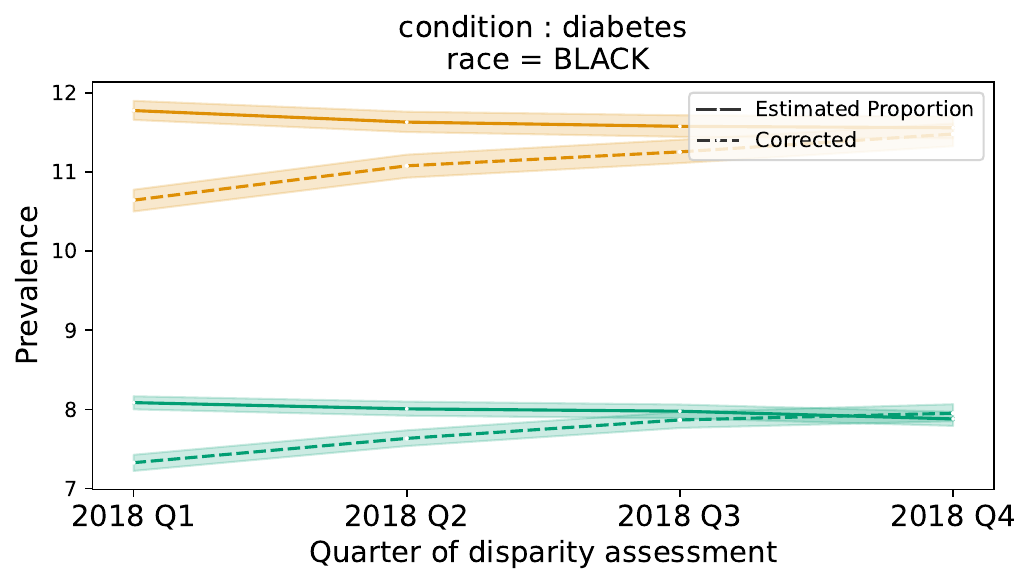}
\end{subfigure}
\begin{subfigure}[t]{0.3\textwidth}
\includegraphics[width=\linewidth]{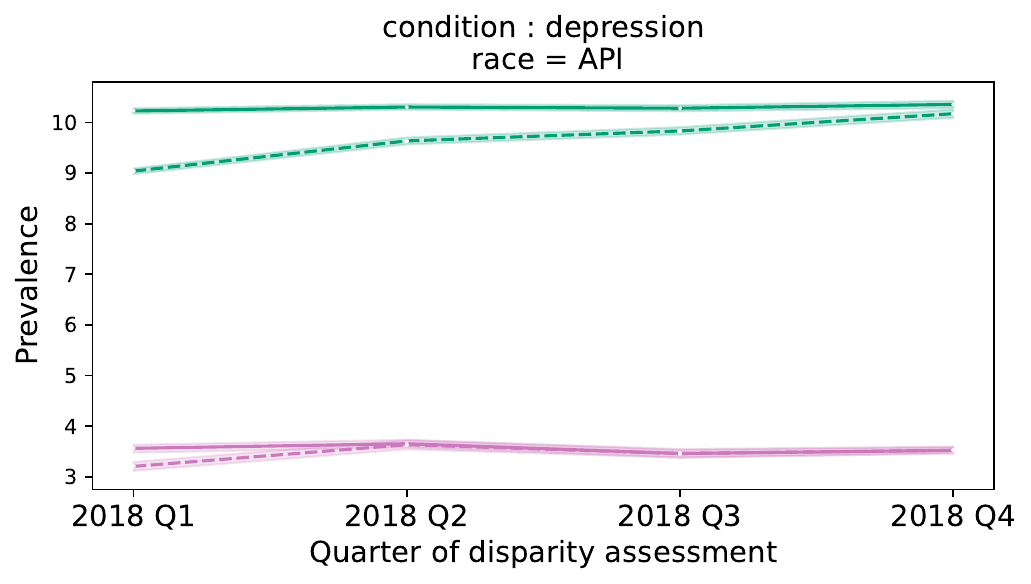}
\end{subfigure}
\begin{subfigure}[t]{0.3\textwidth}
\includegraphics[width=\linewidth]{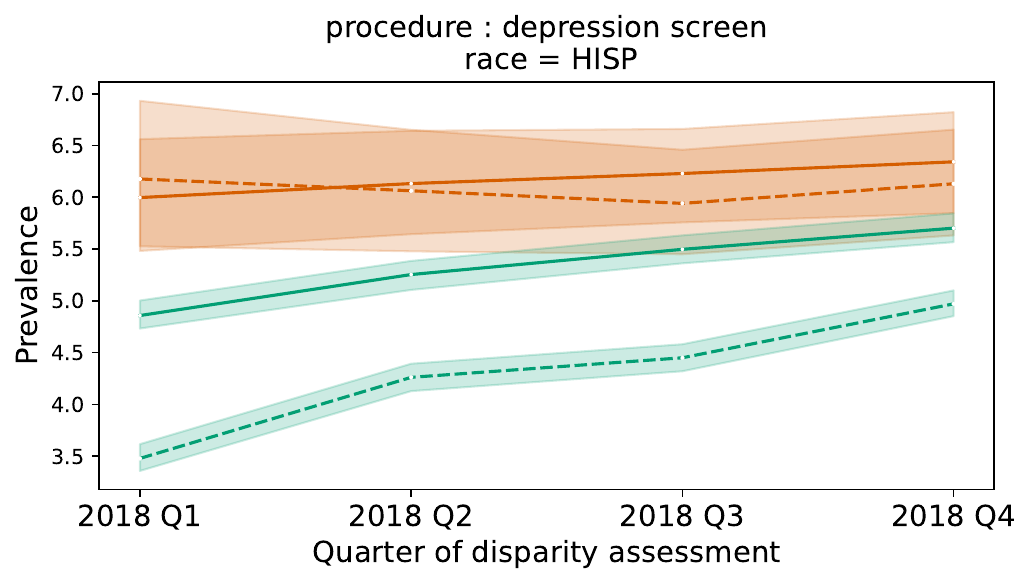}
\end{subfigure}
\caption{\textbf{Comparison of prevalence estimates for different health outcomes at quarterly intervals starting in 2018.} This figure invokes a real-world example of conducting regular disparity estimates given incomplete or delayed information. For each health outcome, we produce $t_{\text{initial}}$ estimates (solid line) and $t_{3Yrs}$ estimates (dashed line) for patients in a minority group (orange, yellow, blue, or purple), in comparison to estimates for White patients (green). We select pairwise comparisons for whichever groups experience the largest average disparity at $t_{3Yrs}$.}
\label{fig:continuous-disparity-monitoring}
\Description{On the x-axis, the quarter of the disparity assessment is shown, ranging from 2018 Q1 to 2018 Q4. On the y-axis, the estimated and corrected prevalence estimates are shown. In general, the disparities shown with the estimated prevalences do not align with the disparities from the corrected prevalences.}
\end{figure*}

The impact of delayed reporting on disparity assessments cannot be disentangled from the real-world setting in which disparity assessments may be used --- \textit{e.g.}, to inform state and local healthcare organizations about any serious health equity gaps and to produce timely interventions. 
In a real-world setting, monitoring would be conducted at regular intervals for different cohorts (\textit{e.g.}, 2018 Q1, 2018 Q2, 2018 Q3, etc.). 
As a result, we now let the clock run beyond a single quarter in 2018 and examine delays in real time. Figure~\ref{fig:continuous-disparity-monitoring} illustrates a full year of disparity assessments, where the evaluation of $t_{\text{initial}}$ (\textit{i.e.}, 2018 Q1 to Q4) and the improved evaluation of $t_{3Yrs}$ appear together. 
Note that at each time step, we only present estimates from $t_{\text{initial}}$ and $t_{3Yrs}$. 
For each outcome we focus on a single pairwise comparison between a White and non-White group, whichever group experiences the largest average disparity at $t_{3Yrs}$ for that outcome over the course of 2018.

This figure underscores that delay can manifest in \textit{every} time step --- a formidable challenge for continuously monitoring disparities. In principle, evidence of disparities would trigger interventions as soon as possible and regular monitoring would subsequently reveal improvements over time. But as long as delays continue to distort ground truth disparity estimates at the same pace as assessments are conducted, decision-makers may need to choose between estimating disparities inaccurately or monitoring health disparities at a slower pace.

\end{document}